\documentclass[article ]{elsarticle}
\usepackage[T1]{fontenc}
\usepackage{lineno,hyperref}
\usepackage{xcolor}
\usepackage{amsmath,amsfonts,amsthm}
\usepackage{graphicx}
\usepackage{epstopdf}
\usepackage{mathtools}
\usepackage{bm}
\modulolinenumbers[5]
\newcommand{\erf}{\mbox{erf}}
\newcommand{\e}{\mbox{e}}
\newcommand{\D}{\mbox{d}}

\journal{Physica A}
\begin{document}

\begin{frontmatter}

\title{Opinion dynamics with emergent collective memory: \\ A society shaped by its own past}

\author[mymainaddress]{Gioia Boschi}
\ead{gioia.boschi@kcl.ac.uk}

\author[mymainaddress]{Chiara Cammarota}
\ead{chiara.cammarota@kcl.ac.uk}

\author[mymainaddress]{Reimer K\"uhn}
\ead{reimer.kuehn@kcl.ac.uk}

\address[mymainaddress]{Mathematics Department, King's College London, Strand, London WC2R 2LS, UK}

\begin{abstract}
In order to understand the development of common orientation of opinions in
the modern world we propose a model of a society described as a large collection of agents that exchange their expressed opinions under the influence of their mutual interactions and external events. In particular we introduce an interaction bias which results in the emergence of a collective memory such that the society is able to store and recall information coming from several external signals. Our model shows how the inner structure of the society and its future reactions are shaped by its own history. 
We provide an analytical explanation of such mechanism and we study the features of external influences with higher impact on the society.
We show the emergent similarity between the reaction of a society modelled in this way and the Hopfield-like mechanism of information retrieval in Neural Networks.
\end{abstract}

\begin{keyword}
\texttt{Opinion dynamics}\sep \texttt{History of Agreement and Disagreement} \sep \texttt{External Information} \sep \texttt{Collective memory} 
\MSC[2010] 00-01\sep  99-00
\end{keyword}

\end{frontmatter}

\section{Introduction}
Contemporary human society lies under the effect of an almost fully accomplished globalisation.
International barriers have become more permeable thanks to the spread of English as universal language. Connections of individuals develop across huge distances throughout the world thanks to the daily use of social media and the immediate coverage nowadays attained by information media. 
These observations naturally support the picture of a global society described as a large collective system involving strongly interacting degrees of freedom, represented by the individuals' actions or opinions.
With this picture in mind, scientists started to study the human society by means of a statistical mechanics approach. 
Early results of this effort resulted in the first opinion dynamics model proposed by a physicist \cite{weidlich1971statistical} and the introduction of the Ising model to study consensus in societies \cite{galam1982sociophysics, galam1991towards} .
An important amount of work followed these first seminal articles, building a literature in which the pressure of the society is modelled by assimilative interactions that mimic the tendency of people to imitation.
The most widely studied models of a society of this kind come from physics, {\it i.e.} the Voter model \cite{clifford1973model} the Ising model with its variants \cite{borghesi2007songs}  already mentioned and the Majority rule model \cite{galam2002minority, martins2013building}. 
Yet, despite many other interesting features, it soon appeared evident that models only based on assimilative interactions describe a society in which full consensus is typically inevitable. To overcome this problem the concept of {\it homophily} \cite{lazarsfeld1954friendship, mcpherson2001birds}, that is the tendency of people to interact more often or with grater intensity with similar others, has been introduced
\cite{axelrod1997dissemination, deffuant2000mixing, hegselmann2002opinion}.
However, the fragmentation of the society into groups with different opinions obtained in models that include both assimilation and homophily was found to be unstable under the introduction of noise \cite{flache2011local, mas2010individualization}. In fact an infinitesimal amount of noise it is seen to irremediably redirect the society to a state of full agreement.
The idea that antagonistic interactions \cite{mas2010individualization, mas2014cultural,  macy2003polarization, flache2011small, sirbu2013opinion, kurmyshev2011dynamics, sznajd2011phase, radillo2009axelrod, martins2010mass, baldassarri2007dynamics} and xenophobia (the phenomenon for which the larger the dissimilarity between two interacting individuals, the more they evaluate each other negatively) \cite{macy2003polarization, flache2011small, baldassarri2007dynamics,  mark2003culture} should be taken into account to resolve this issue has been developed only fairly recently.
In the present work we build up on these observations and put them in conjunction with the traditional approach that assumes imitation tendencies between individuals.
We propose a model for opinion formation based on the rule that individuals in agreement with each other tend to reinforce their mutual positive influence, while individuals in disagreement will develop an antagonistic relation based on the mistrust towards one another's view. These tendencies will be encoded in an interaction term that for each pair of individuals reflects the history of their agreement or disagreement at previous times. Past relevant interactions are only those that lie within the finite range of agent's memory.\\

We note that the dynamics of the model we propose is strongly reminiscent of the dynamics of graded response neural networks \cite{hopfield1984neurons, kuhn1991statistical, amit1987statistical} which have been used to describe associative memory. Indeed, the model discussed here will, under suitable conditions, develop interactions of the Hopfield type \cite{hopfield1982neural}. 
Models of society including Hopfield-like interactions have been already used in social sciences in a few occasions \cite{macy2003polarization,flache2011small} to study consensus formation and opinion polarization.
In these works Hopfield interactions are introduced and studied in conjunction with a number of other elements like individual flexibility, broad-mindedness, and open-mindedness \cite{macy2003polarization} or in more complicated network structures \cite{flache2011small}.
Moreover in all these cases the focus of the study was on the stationary state reached by a society with fixed interpersonal interactions and how it is approached from a random initial condition. 
Our contribution will instead focus on a model society whose internal interactions develop starting from historical interpersonal relationships. Yet, it will, under suitable conditions, spontaneously turn out to closely resemble a model society exhibiting Hopfield-like couplings.
We will develop a detailed analysis of the way the Hopfield-like interactions develop and analyze the resulting collective behaviour of the system.
To this aim we will focus on a society that is constantly under the influence of new external events and we will pay particular attention to the reaction of the society to world-wide news, modelled as external fields applied to it \cite{carletti2006make, gargiulo2008saturation}.
We will study whether the society does develop self-maintained collective memories of certain news, and will therefore be irremediably shaped by them. In literature the concept of collective memory was first introduced in \cite{halbwachs1992collective} and only recently scholars have focused on how collective memory is influenced by media  \cite{kligler2014setting, garcia2017memory}. In our model the possibility for external events to impress the society will depend on a number of parameters including the extent of the news influence on single individuals and how frequently they are impacting the society. In particular our model shows that strongly impacting news, or even just very frequent news, can change the internal structure of the society in a drastic way and will determine its non linear collective response to future external influences.
\noindent The structure of the paper is as follows.
In Section \ref{model} we will introduce the model and its main features, before entering into the description of different scenarios corresponding to different kinds of external information. The results are presented in Section \ref{results} and organized in the following sections starting from the simplest scenario to more complicated ones. 
Given the growing complexity of the problems studied, not all
the cases considered can be fully solved analytically. 
For each choice of the external stimuli we first derive all the analytic predictions we have access to, then we complete the picture by showing simulation results.
\section{The model}
\label{model}
 The society that we consider is composed of a set of $N$ agents, each potentially connected to all the other agents. With agent $i$ we associate a continuous preference field $u_i$.
 Expressed opinions are given by nonlinear functions $g(u_i)$ of the preference field $u_i$ of agent $i$. We will take $g(u_i)$ to be of sigmoid form, implying that expressed opinions remain bounded. 
We will take the stochastic dynamics of the system to be of the form: 
\begin{eqnarray}\label{maineq}
\dot{u_i}=-u_i+I_i+\sum_{j\neq i}^N J_{ij}g_j+\eta_i\ ,
\end{eqnarray}
where we use the abbreviation $g_j= g(u_j)$ and we dropped time dependencies, which in general pertain to all variables in the equation. Here $J_{ij} >0 $ represents a mutually supportive interaction between the individuals $i$ and $j$ while $J_{ij}<0$ indicates an antagonistic interaction between the same agents. The quantity $I_i$ represents the mass media information as perceived by individuals, while $-u_i$ is a mean reversion term which entails that in absence of external influences the preference field of each agent will fluctuate around zero.
The last term $\eta_i$ is a white noise with Gaussian distribution with zero mean and finite variance $\langle \eta_i(t)\eta_j(t')\rangle = \sigma^2 \delta_{ij} \delta(t-t')$.
The combined effect of individually perceived external sources of information, $I_i$, and the interaction with others agents' expressed opinions, $g_j$, may act to drive the preference field of an agent away from zero, and thus may favour the development of specific orientations. 
The heterogeneous external information $I_i$ is defined as $I_i=I_0 \xi_i$ separating the strength $I_0$ of the signal from the variables $\xi_i$ encoding the local variability. Note that such variability might be genuine or arise as a result of individually variable perception of an underling uniform message (which may be caused by idiosyncratic interpretations).
We simply assume that $\xi_i$ can only be either $+1$ or $-1$. \\
 Finally and most importantly the tendency of each person to agree or disagree with others is based in our model on the memory of past history of agreement and disagreement with them. People which have a history of agreement in the past, will be more likely to agree also in the future, and an analogous result holds for disagreement. 
 This feature represents the key ingredient of our model.
 More specifically we consider the recent history of agreement or disagreement to have a larger weight than the distant past to take into account how vivid the experience of past interactions is. We will assume an exponentially weighted memory and take interactions between agents at time $t$ to be given by  
\begin{eqnarray}\label{eqJs}
J_{ij}(t)= \frac{J_0\cdot \gamma}{N} \int_{0}^{t}\mbox{d}s \ g_i(s)g_j(s) e^{-\gamma(t-s)} 
 \end{eqnarray}  
for some $J_0>0$, assuming for simplicity the time scale $\tau_\gamma=1/\gamma$ of the memory to be uniform across agents.
 The normalization factor $1/N$ is introduced so that the interaction with other agents is not overwhelmingly dominant, but remains comparable to the influence of external sources of information in the large $N$ limit.\\
 In this way we have a society that uses the past history to interpret any instantaneous inputs that it receives. In particular agreement (disagreement) will be perceived if the value of $g_i(s)g_j(s)$ will be continuously positive (negative) on the time scale $\tau_\gamma$ and will bias agents $i$ and $j$ toward future agreement (disagreement). 
 When the history of past interaction is instead characterized by an alternation of agreement and disagreement periods the agents will tend to be neutral with each other, $J_{ij}\sim0$.
 This memory effect is particularly important when studying the influence of the external information on the agents' opinions.
Note that the memory that appears in our model, being a memory of past relations, must not be confused with the memory of past actions or opinions of single agents that has been more often considered in the literature of social behaviour \cite{dall2007effective, noah1998beyond, jkedrzejewski2018impact}. \\
\subsection{Agents' interactions {\it vs} Hopfield couplings}
\label{Hopfield_couplings}
The main ingredient of our model is the memory of past interactions, which is associated with the time scale $\tau_\gamma$. The model also contemplates a second time scale which is the relaxation time of the individual preference $u_i$. The latter has been set equal to 1, without loss of generality as in general all the other parameters and time can be expressed in terms of its unit. 
A third time scale $\Delta_0$ should be also considered. It is the one associated with the duration of the exposure of the society to external stimuli.
We will typically focus on the regime $\tau_{\gamma} \gg\Delta_0\gg 1$ corresponding to a fast adaptation of $\mathbf{u}$ to eventual external stimuli and a slow memory decay which will be responsible of the storing of previous opinion configurations in the memory of interpersonal relations. This process will describe how the whole society can be {\it shaped} by its past by {\it learning} from patterns of opinions produced by sustained signals or series of repeated external stimuli. 
Among the different scenarios studied, we will describe the case of the arrival of different external stimuli represented by a local field changing in time $\bm{I}(t)=I_0\bm{\xi}^{\mu(t)}$. In this expression $p$ different random choices of ${\bm \xi}^{\mu(t)}=(\xi_i^{\mu(t)}) \in  \{\pm 1\}^N$ are considered, one for each integer value in $\{1...p\}$ that $\mu(t)$ assumes, each of them for a time $\Delta_0$. Each of these random vectors represents the perceived piece of news that influence the society during the time $\Delta_0$ and is later substituted by a different piece of news. Under the effect of such external influences we expect that the society will likely develop interactions comparable to the classic Hopfield couplings \cite{hopfield1982neural} defined from a collection of $p$ random patterns ${\bm \xi}^{\mu}$, albeit rescaled by a factor $1/p$, so in the long time limit we expect
\begin{equation}\label{10}
J_{ij} \simeq J_{ij}^{\mbox{H}}/p=\frac{1}{Np}\sum_{\mu=1}^p\xi_{i}^\mu\xi_{j}^\mu\ .
\end{equation}
In fact, if each strong signal clamps the expressed opinions towards the positions that it suggests ($\mathbf{g}(t)=\bm{\xi}^{\mu(t)}$) for a time $\Delta_0\ll\tau_\gamma$, and the sequence $\mu(t)$ of the signals' appearances is repeated many times in $\tau_{\gamma}$, 
the average over the past history in Eq.(\ref{eqJs}) can be approximated by an average over the product of $\xi_{i}^\mu$ and $\xi_{j}^\mu$ as by definition of Hopfield couplings. 
 In section \ref{sec:per_ext_sig} we will discuss the similarity between the generated interpersonal couplings in our modeled society and Hopfield couplings in more detail.
For the moment it is interesting to note that, despite the similarity that emerges at first sight, the couplings in the Hopfield model \cite{hopfield1982neural, hopfield1984neurons} were taken to be fixed from the start, whereas in the present case they evolve dynamically under the influence of external signals and internal dynamics.  It is only in the case of the special scenario described above that we will see the emergence of Hopfield type couplings. 

\subsection{Numerical details}
While a few of the simplest scenarios investigated in the present paper can be studied analytically, we are in many cases forced to use numerical simulations. To perform these, we note that the couplings of Eq. \eqref{eqJs} satisfy a dynamical evolution equations:
\begin{equation}\label{dynrule}
\dot J_{ij}(t)= \gamma \left[\frac{J_0}{N}g_i(t)g_j(t) - J_{ij}(t)\right]\ .
\end{equation}
We use Euler integration to integrate Eq. \eqref{maineq} and Eq. \eqref{dynrule}; we found a step size
$\D t=0.1$ (with $\D \eta_i(t)=\sigma \sqrt{dt}$) sufficiently small for our purposes.
There are many parameters involved in the simulations, some are fixed in all the cases, some vary. The fixed parameters are listed here: we choose $N=100$ for the number of agents and $p=3$ for the number of different external signals. Although these numbers are small, they produce results that are representative of the $N\to \infty$ limit with $p\ll N$ (as we have checked by using other $(p,N)$ combinations). Throughout this paper we used a low noise level, $\sigma^2=0.01$, to ensure that non-trivial collective states can emerge. All simulations start with random initial conditions $u_i \sim \mathcal{N}(0, \sigma^2/2)$ which would be the equilibrium distribution in a non-interacting system without external signal. 
The other important parameters that change from case to case are the time length of each external signal $\Delta_0$, the amplitude $I_0$ of the polarizing signal (apart from a few exceptions taken to be $I_0=50$), the strength of the interactions $J_0$ and the time scale $\tau_{\gamma}$ of the memory of past interactions.\\

\section{Results Overview}
\label{results}
Our model is constructed on the simple assumption that the mutual interactions between agents depend  on their past history of agreement or disagreement. Our main result is that this creates a mechanism that allows a society to develop a collective memory of its past experiences. 
To study this mechanism in more detail, we analyze different protocols of external influences that first trigger the different individuals' opinions. The different scenarios presented range from simple situations to more complex and realistic ones and are listed below together with a list of the results obtained in each case. \\ 
\begin{enumerate}
    \item The external information $\bm{I}$ is heterogeneous but constant in time. We are able to treat the system analytically and predict its long time behaviour. Already in this simple setting the presence of the signal changes the way people interact and determines their future behaviour. 
    \item The signal consists of a sequence of different (random) patterns, repeatedly presented in a cyclic fashion. In the long run this creates a stable matrix of interactions between the agents which can be predicted analytically. 
    By means of this interaction matrix  the society develops a memory of the opinion patterns presented previously     and it is found to be able to recall each of them. 

    \item A sequence of external stimuli is repeatedly presented in a cyclic fashion as before. However there are gaps between the presentation of successive signals where the society is not exposed to an external stimulus and follows its own internal dynamics. In this scenario a critical ratio of patterns duration and length of the gaps without pattern presentation must be exceeded for the system to develop persistent memory of the signal. We provide an analytic treatment to predict this critical ratio. 
    Interestingly we also found that in the case of high-impact news that influence the society very frequently, the presentation time needed for the news to be memorized by the society is unexpectedly small.
\end{enumerate}
The study of even more realistic situations such as sequences of external influences with different impact on the society and random appearance in time will be considered in a follow up paper.
\section{Learning from a persistent external signal}
\label{one_pattern}
In this section we start illustrating the behaviour of the society described by our model in a simple setting. Here we study its reaction to an external information persistent in time and described by a signal $I_i=I_0 \xi_i$ where $I_0$ is its strength and $\xi_i$ is a random variable taking values in $\{\pm 1\}$, which represents the way in which the agent $i$ perceives it. Note that the uniform perception $I_i=I_0 \ \ \forall i$ is a particular case of what discussed here. Our aim is to understand how the society reacts to this signal and how the memory of the opinions induced by it develops in this simple case before moving to more complicated settings. In order to do this, we will study the evolution in time of the agents' preference field $u_i$ and of interaction couplings $J_{ij}$ between agents. 
Even in the simple case of a constant signal, solving the equation for $u_i$ is not a trivial task, mainly because of the dependence of the couplings on the expressed opinions.
We will see that the presence of the signal will induce the agents to change their opinion, consequently modifying the relationship of agreement and disagreement between them. As a results of this change, the couplings $J_{ij}$, which are null at the beginning of the dynamics, will start to evolve and establish the new interactions between the agents. This process allows the society to learn the opinion pattern $\bm{\xi}$ and to collectively retrieve it in the future. \\
We can write down a formal solution of Eq. \eqref{maineq} with couplings defined in Eq. \eqref{eqJs}: 
\begin{eqnarray}\label{u1}
 u_i(t)&=& u_i(0)e^{-t}+ J_0 \int_{0}^{t}\D s \left[ U_i(s)+\eta_i(s) \right] e^{-(t-s)}+I_0\xi_i (1- e^{-t}) \ ,
\end{eqnarray}
in which 
\begin{equation}\label{U1}
U_i(s)=\gamma \int_0^s\D s' \e^{-\gamma(s-s')}g_i(s')q(s,s')\ ,
\end{equation}
with 
\begin{equation}
    q(s,s')=\frac{1}{N}\sum_jg_j(s)g_j(s')\ .
\end{equation}
We note that, by appeal to the law of large numbers, the correlator $q(s,s')$ will be non-random in the large $N$ limit. Noting further that, for $\gamma \ll 1$, the function $U_i(s)$ are very slowly varying funcions of $s$, we see that:
\begin{equation}\label{U2}
    \int_0^t \D s U_i(s)\e^{-(t-s)} \simeq U_i(t) \int_0^t \D s \e^{-(t-s)} = U_i(t)(1-\e^{-t})\ .
\end{equation}
To proceed we adopt one further approximation to replace $U_i(t)$ in Eq. \eqref{U2} by its noise average (indicated by $\langle \cdot \rangle$), which is tantamount to replacing the $g_i(s')$ in Eq. \eqref{U1} by their noise average. While this replacement leads to an underestimation of the noise contribution in the evolution of the $u_i(t)$, we found that the effect remains small in the parameter ranges considered. With this approximations the solution for $u_i$ becomes
\begin{equation}\label{u3}
 u_i= u_i(0)\e^{-t}+ J_0 \langle U_i(t)\rangle (1-\e^{-t}) + \int_{0}^{\infty}\eta_i(\tau)e^{-\tau}\D \tau+ I_0\xi_i(1-\e^{-t}) \ .
\end{equation}
This means that $u_i$ will be a Gaussian process with mean
\begin{equation}
    \langle u_i \rangle = u_i(0)\e^{-t}+ J_0 \langle U_i(t)\rangle (1-\e^{-t}) + I_0\xi_i(1-\e^{-t})\ ,
\end{equation}
covariance
\begin{eqnarray}
C(t,t')&=&\langle(u_i(t)-\langle u_i\rangle \nonumber (t))(u_i(t')-\langle u_i\rangle (t'))\rangle \\
 &=&\int_{0}^{t}\int_{0}^{t'}\D s \D s' \langle \eta_i(s)\eta_i(s')\rangle e^{-(t-s)}e^{-(t'-s')} \nonumber\\ 
&=&\frac{\sigma^2}{2} \left(\e^{-|t-t'|}-\e^{-(t+t')}\right)\ ,
\end{eqnarray}
and variance 
\begin{equation}
\sigma_u^2(t)=C(t,t)=\frac{\sigma^2}{2}(1-\e^{-2t})\ .
\end{equation}
We are interested in studying the system in the long time limit $t, t' \to \infty$, where the $u_i$-process become stationary, with the covariance of the $u_i$ only depending on the time difference $\tau=|t-t'|=O(1)$, i.e. $C(t,t')=C(\tau)$, so we will have
\begin{equation}
\lim_{t \to \infty} \langle U_i(t)\rangle = \gamma \langle g_i \rangle \tilde{q}(\gamma)  
\end{equation}
where $\tilde{q}(\gamma)$ is the Laplace transform of $q(t-s')$. 
The mean of $u_i$, its covariance and variance will thus become
\begin{eqnarray}
\langle u_i \rangle &=& \gamma \langle g_i \rangle \tilde{q}(\gamma) J_0  + I_0\xi_i \label{mean_u}\\
C(\tau) &=&\frac{\sigma^2}{2} \e^{-\tau} \label{24} \\
\sigma^2_u &=&C(0)=\frac{\sigma^2}{2} \label{sigma_u}\ .
\end{eqnarray}
Now we realize that $\langle u_i \rangle=\xi_i\langle u \rangle$ and $\langle g_i \rangle = \langle g(\langle u_i \rangle + \sigma_u \zeta_i)\rangle_\zeta = \xi_i \langle g \rangle$, with $\langle \cdot \rangle_\zeta$ being the average taken over $\zeta \sim \mathcal{N}(0,1)$, are a consistent solution of Eq. \eqref{mean_u}. This allows us to take out the dependencies on $i$ from the equations.\\
If we now choose the expressed opinions to be the error functions of the preference fields, with $g_i=\erf(u_i)$, we can exploit the properties of the error function to obtain a self consistency equation for $\langle u \rangle$. We will have
\begin{equation}
\label{g}
    \langle g\rangle=\erf\left(\frac{\langle u\rangle}{\sqrt{1+2\sigma_u^2}}\right)
\end{equation}
and 
\begin{equation}
\label{q_tau}
q(\tau)=\left\langle\mbox{erf}\left(\langle u\rangle_\infty+\sigma_{u}x\right)\mbox{erf}\left(\frac{\langle u\rangle _\infty+\rho(\tau)\sigma_{u}x}{\sqrt{1+2(1-\rho^2(\tau))\sigma_{u}^2}}\right)\right \rangle_x\ ,
\end{equation}
where $\langle\cdot\rangle_{x}$ stands for an average over a Normal random variable $x$ and the correlation coefficient $\rho(\tau)$ is 
\begin{equation}
\rho(\tau)=\frac{C(\tau)}{\sigma_u^2}=e^{-|\tau|}\ .
\end{equation}
(See \cite{anand2018structural} for further details on these last passages from a similar computation in a different context.) \\
Gathering these results together we obtain a closed system of equations that describe the preference field of the agents in our society under the effect of a constant external signal in the stationary regime:
\begin{eqnarray}
\nonumber
\langle u\rangle &=& I_0 + J_0 \gamma \mbox{erf}\left(\frac{\langle u\rangle}{\sqrt{1+2\sigma_u^2}}\right)\tilde{q}(\gamma)\ ,\\ \nonumber
\sigma_u^2&=&\frac{\sigma^2}{2}\ ,\\ \nonumber
\tilde{q}(\gamma)&=&\int_{0}^{\infty}\D \tau q(\tau)e^{-\gamma \tau} \ ,\\ \nonumber
q(\tau)&=&\left\langle\mbox{erf}\left(\langle u\rangle+\sigma_{u}x\right)\mbox{erf}\left(\frac{\langle u\rangle+\rho(\tau)\sigma_{u}x}{\sqrt{1+2(1-\rho^2(\tau))\sigma_{u}^2}}\right)\right\rangle_x\ .
\end{eqnarray}
In  order  to  understand  to  what  extent  the  external  field  has  influenced  the
society we will focus on the overlap of the asymptotic system state with the pattern $\bm{\xi}$:
\begin{equation}
\label{m}
    m(t) = \frac{1}{N}\sum_i \xi_i g_i(t) \ .
\end{equation}
This quantity measures whether the expressed opinions in the society are similar to those induced by the signal $\bm{\xi}$, {\it i.e.} $m\sim O(1)$, or not. The overlap is self-averaging in the $N \to \infty$ limit and its value at late times, $t \to \infty$, is given by
\begin{equation}\label{self_m1}
    m = \frac{1}{N}\sum_i \xi_i \langle g_i\rangle = \langle g \rangle \ ,
\end{equation}
which therefore satisfy the following equation
\begin{equation}\label{self_m2}
m = \erf\left(\frac{ \langle u \rangle }{\sqrt{1+2\sigma_{u}^2}}\right)=\erf\left(\frac{ I_0 + J_0 \gamma  \tilde{q}(\gamma)m}{\sqrt{1+2\sigma_{u}^2}}\right)\ .
\end{equation}
We chose to write the last equation in a self consistent form to show that under certain conditions we can expect a non zero value of $ m$ even when the signal is finally removed, $I_0=0$. 
Indeed, for sufficiently large values of the product $J_0\tilde{q}(\gamma)$ in comparison with the noise contribution quantified by $\sigma$, the equation admits a non zero solution. For example, if the society is exposed to a signal $I_0=1$ for long times, given a $J_0=6$ and $\gamma=10^{-3}$ its overlap will be close to 1 and will remain close to 1 when the signal is removed (we will see that in the same conditions this value will match the results obtained with a simulated dynamics).
In other words the society is potentially able to remember the opinions induced by the signal even when it is removed, after having been exposed to such signal for sufficiently long time. 
To shed more light on this mechanism we now focus on the evolution of the couplings and the value they reach in the stationary regime for a society described by a choice of the parameters that admits a non zero solution to Eq.\eqref{self_m2}.
\begin{figure}
    \centering
    \includegraphics[scale=0.6]{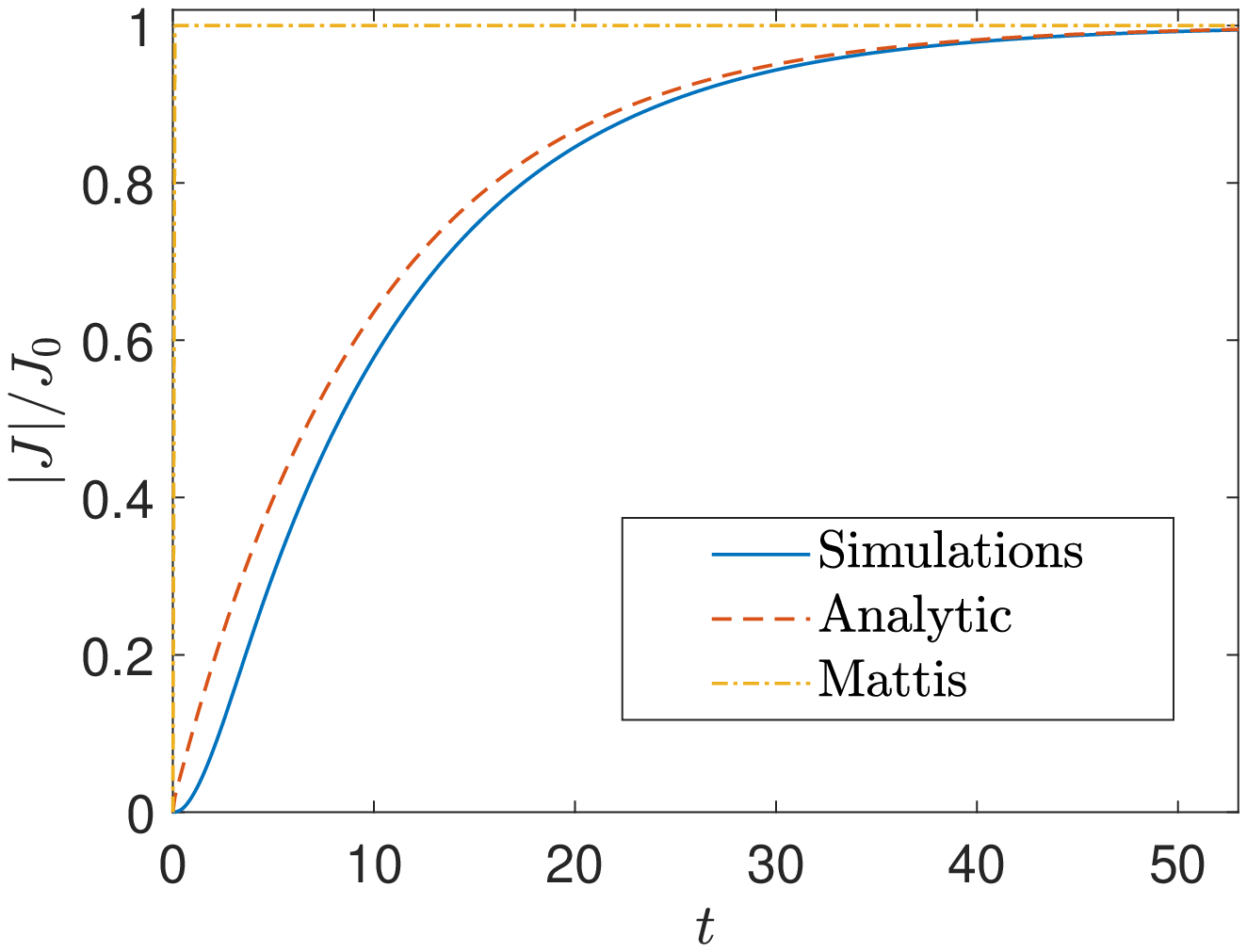}
    \includegraphics[scale=0.6]{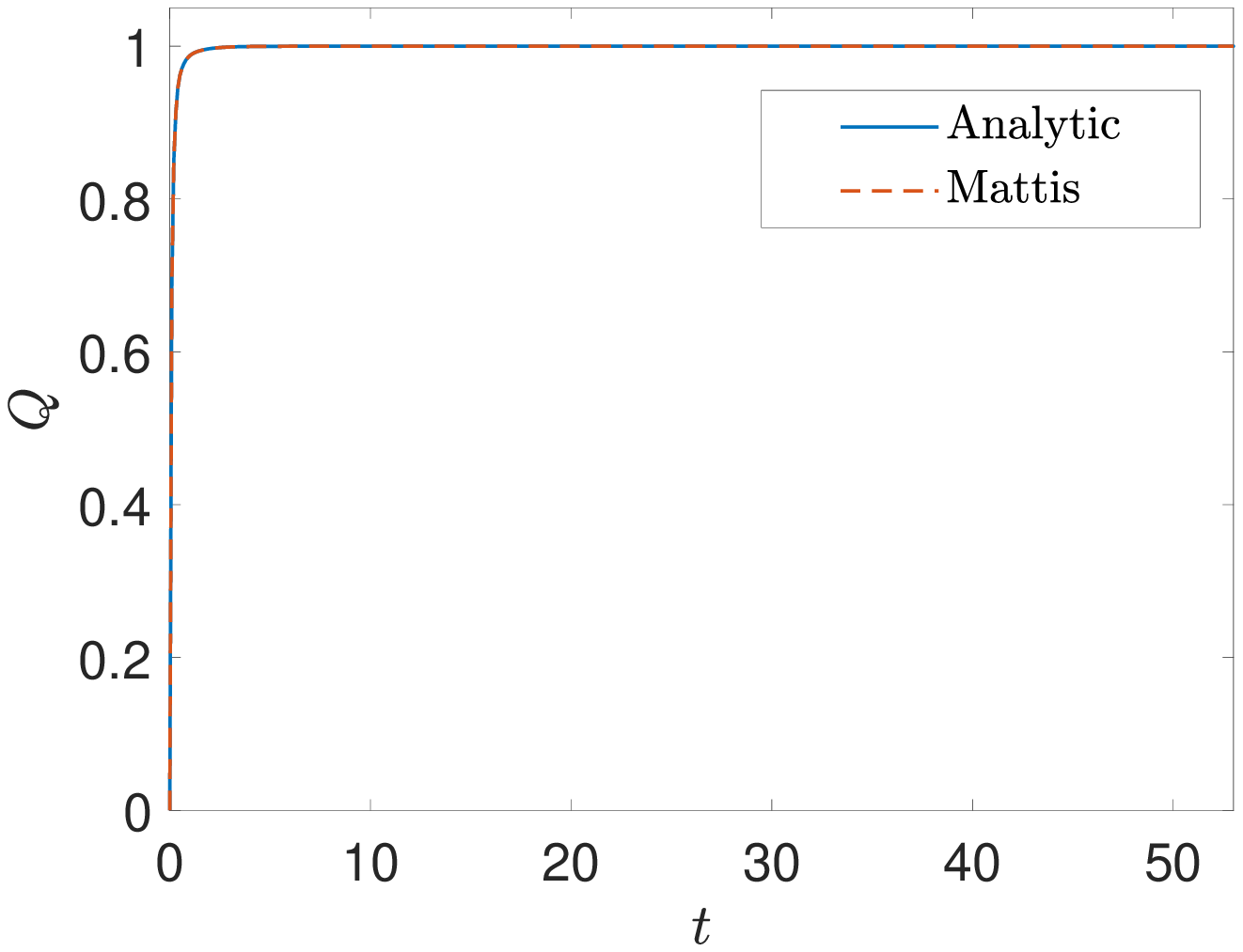}
    \caption{Simulated dynamics with a persistent external signal. Upper panel: results from simulations for $|J|/J_0$ are compared with the analytical prediction (Eq. \eqref{J_analytic}) and are seen to approach the asymptotic Mattis couplings (Eq. \eqref{mattis}). Lower panel: the overlap $Q$ between the simulated and the analytical $J$ is compared with the overlap between the simulated $J$ and the Mattis couplings. In these simulations $J_0=6$, $I_0=1$ and $\gamma=0.1$.}
    \label{one_pattern_m}
\end{figure}
 The upper panel of Fig. \ref{one_pattern_m} shows how couplings evolve in a numerical simulation starting from a society without interactions between agents. The continuous curve shows the norm of the $J_{ij}$ matrix defined as
\begin{equation}
\label{absJ}
   |J|=\sqrt{\sum_{ij}J_{ij}^2}\ .
\end{equation}
For sufficiently large $I_0$, a simple analytic argument allows us to give an accurate prediction of this growth. In the presence of a signal $I_i=I_0\xi_i$, preference fields rapidly orient towards the signal with large absolute values of $u_i$ and $g_i$ will therefore soon be approximately equal to $\xi_i$. By using this information we can integrate Eq. (\ref{eqJs}) to obtain:
 \begin{eqnarray}\label{J_analytic}
J_{ij}(t)\approx \frac{J_0\cdot \gamma}{N} \int_{0}^{t}\mbox{d}s \ \xi_i\xi_j e^{-\gamma(t-s)}= \frac{J_0}{N} \xi_i\xi_j (1-e^{-\gamma t}) \ ,
 \end{eqnarray} 
where the memory time scale appears explicitly. 
The average absolute value of this analytic prediction is represented also in units of $J_0$ by the dashed line in the upper panel of Fig.\ref{one_pattern_m} and nicely superimpose with the numerical results. As the system will quickly approach a stationary state, aligned with the external signal, interactions further stabilizing that the very same stationary state will be established after a time set by the memory time scale, chosen to be $\tau_\gamma= 10$ in that simulation. \\
Note that the sign of the predicted couplings is also peculiar. 
For long times the $J_{ij}$ approach the Mattis couplings \cite{mattis1976solvable}:
\begin{eqnarray}\label{mattis}
  \lim_{t\rightarrow\infty}J_{ij}(t)= \frac{J_0}{N} \xi_i\xi_j \equiv J^{M}_{ij}
\end{eqnarray} 
equivalent to the Hopfield couplings in Eq.(\ref{10}) for p=1. 
It is well known that in a system with pairwise interactions given by Mattis couplings (Eq.\eqref{mattis})  with sufficiently large amplitude, the variables spontaneously align in the directions defined by $\bm{\xi}$.
In our modelled society this would correspond to the fact that the opinion pattern can be spontaneously retrieved because the corresponding Mattis couplings have been formed as a consequence of the memory of the sustained past opinion patterns induced by the exposure to the signal $I_i=I_0\xi_i$. 
To verify that this is the case we define the correlation
\begin{eqnarray}
Q=\frac{\sum_{ij}J_{ij}J_{ij}'}{\sqrt{\sum_{ij} J_{ij}^{2}\sum_{ij} {J'}_{ij}^{2}}} \label{2}
\end{eqnarray}
that reveals the degree of alignment between two sets of couplings $J=(J_{ij})$ and $J'=(J'_{ij})$, with $Q=1$ implying perfect alignment and $Q=0$ the absence of any correlations.
In the lower panel of Fig.\ref{one_pattern_m} we exhibit the correlation between couplings observed in a simulation with both the analytic prediction (Eq. \eqref{J_analytic}) and the asymptotic Mattis couplings (Eq. \eqref{mattis}). It shows that interactions are very quickly perfectly correlated with both the analytically predicted couplings and the Mattis couplings, although their norm is initially smaller than $|J^M|$ (see comparison with the absolute value of $J^M/J_0$ in the upper panel of Fig.\ref{one_pattern_m}).
In conclusion, the memory of interpersonal relations developed in response of an external stimulus $\bm{\xi}$ produces in the society interactions of increasing strength, which are very quickly aligned with Mattis couplings corresponding to the $\bm{\xi}$ itself. \\
For large enough strength of the couplings we can expect that the society will spontaneously polarize along $\bm{\xi}$ autonomously sustaining its memory even after the external signal is gone. 
To give evidence of this interesting phenomenon we studied different dynamics in which the external signal $\bm{\xi}$ is switched on at $t=0$ and removed at a time $t_r$.
We then focus on the \textit{asymptotic} average overlap for
\begin{enumerate}
    \item dynamics in which couplings keep evolving after $t_r$ (evolving $J_{ij}$)
    \item dynamics in which couplings are fixed to the value they had at $t_r$ (frozen $J_{ij}$).
\end{enumerate} 
\begin{figure}
    \centering \includegraphics[scale=0.6]{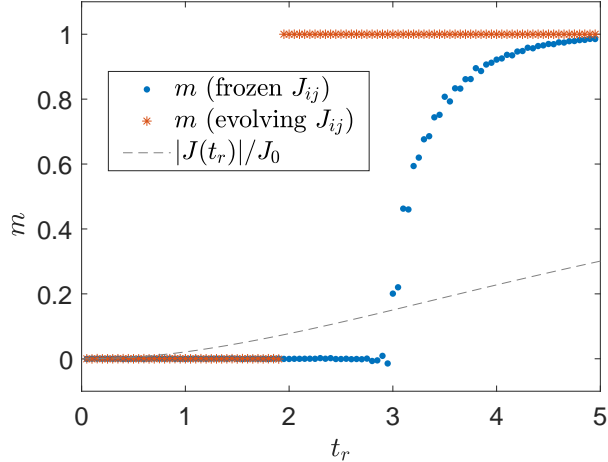}
    \caption{Simulated dynamics with a persistent external signal. The figure shows the value of $m$ at stationarity as a function of the time $t_r$ at which the signal is removed, for two kind of dynamics: 1) dynamics with evolving couplings after $t_r$ 2) dynamics with frozen couplings after $t_r$. The dashed line represents the amplitude of couplings norm $|J(t_r)|$ at $t_r$. 
    The parameters used are the same as Fig. \ref{one_pattern_m}.}
    \label{one_pattern_m2}
\end{figure}
The average overlap for both versions of the dynamics has been calculated as the average of the simulated overlap at stationarity over the last 1000 units of time and plotted in Fig. \ref{one_pattern_m2} as a function of $t_r$. \\
The dynamics with frozen $J_{ij}$ shows at which $t_r$ the couplings have grown large enough for a spontaneous $m$ to arise. %in a continuous way. 
We note here that this overlap $m$ corresponds to the solution of the self-consistent Eq. \eqref{self_m2} derived for the stationary solution once the amplitude of the asymptotic couplings $J_0 \tilde{q}(\gamma)$ is replaced by the amplitude of the frozen couplings $|J(t_r)|$ (also reported in Fig. \ref{one_pattern_m2}) and $I_0=0$.
According to this equation the critical value of the strength of Mattis type couplings needed for the model to exhibit spontaneous order with $m>0$ is $|J_c|=0.148 J_0$ and it corresponds to the $|J|=(0.148 \pm 0.003)J_0$ reached at the minimum $t_r$ where $m>0$ in simulations with frozen $J_{ij}$.
We will come back to this minimum time $t_{r}$ 
in a different context in Section \ref{sec:Learning with signal removal}. \\
Interestingly the dynamics with evolving $J_{ij}$ is instead characterized by an \textit{asymptotic} spontaneous overlap arising in a discontinuous way even before that minimum $t_r$. 
Indeed, at variance with the frozen case, the evolving couplings will continue to grow after $t_r$ even in absence of the external signal as a result of the interactions embedded in the system. If this residual reinforcement of the couplings is large enough a positive feedback loop will occur between the evolving couplings and the degree of order supported by the interactions currently established in the society. The degree of order will be strengthened by strong interactions, which will in turn grow further due to a higher degree of persistence of the expressed opinion pattern.
As soon as this self-sustained mechanism takes place it leads to a strongly polarized society represented by the large value of $m $ in Fig. \ref{one_pattern_m2}.\\
In the following sections we will study more complicated cases in which the external signal is composed of a sequence of different stimuli both with and without interposed periods of complete absence of stimuli. 
The understanding of the system's reaction to a single external stimulus gained in the present section as well as the quantities introduced here will be used in the next sections to understand if and how the society is able to learn and subsequently retrieve the different opinion patterns to which it has been exposed in the past.

\section{Periodic external signal}
\label{sec:per_ext_sig}
 As a second step in the exploration of our society's behaviour we will change the external information structure, passing from a constant signal $I_0\bm{\xi}$ to a signal that changes in time. The aim is mimicking a sequence of different news or events, labeled by the index $\mu \in \{1, \dots, p\}$, to which the society is exposed and reacts.
 As before, the variables $\bm{\xi}^\mu$ represent the way in which the population reacts to a single piece of information, and are modelled as random variables with entries $\xi_i^\mu$  $\in \{\pm 1\}$.
 The different contributions $I^\mu=I_0\bm{\xi}^\mu$ will appear in an ordered sequence, each  switched on for a time span $\Delta_0$ before being substituted by the following piece of news in the sequence. 
 This process is repeated in a cyclic fashion. The resulting expression of the signal is thus
 \begin{equation}\label{18}
 I_i(t)= I_0 \xi_i^{\mu(t)}\ ,
 \end{equation}
 with
 \begin{equation}\label{32}
 \mu(t)= 1+\left \lfloor{\frac{t}{\Delta_0}}\right \rfloor \hspace{-0.3cm}\mod{p}\ ,
 \end{equation}
where $\left \lfloor{\cdot}\right \rfloor$ is the notation for the integer part. This series of repetitive signals represents a simple but effective way to model series of events that are repeatedly appearing in television or newspapers.\\
As introduced in section \ref{Hopfield_couplings}, we expect that in the long run the exposure of the society to a sequence of signals defined in Eq. \eqref{18} and \eqref{32} will result  in the development of couplings that are similar to the Hopfield couplings (Eq. (\ref{10})).
Indeed as before, given a signal with large amplitude $I_0$, we can assume that the opinions $g_i$ become rapidly equal to the opinion pattern proposed every time we have a signal spike, so we have that $\bm{g}(t)=\bm{\xi}^{\mu(t)}$. For this assumption to provide an accurate approximation of the full dynamics, we also assume that $\Delta_0\gg 1$ allowing us indeed to neglect transient behaviour after the switches of the external signal.
Using this approximation we can calculate (see Appendix A for details) the couplings $J_{ij}$ developed in the society at times $t$, which are multiples of the presentation time $p\Delta_0$: 
\begin{eqnarray} 
\label{J_signal_nodrop_2} J_{ij} = \frac{J_0}{N}  (e^{\gamma \Delta_0}-1) \frac{(e^{-\gamma t}-1)}{(1-e^{\gamma \Delta_0 p})}\sum_{\mu=1}^{p}\xi_i^\mu\xi_j^\mu e^{(\mu-1)\Delta_0\gamma} 
\end{eqnarray}
In the long time limit $t \to \infty$ (thus $N_p\to \infty$) this tends to
\begin{eqnarray}
\label{J_inf}
\lim_{N_p\to \infty} J_{ij}(t=N_p\Delta_0 p) = J_{ij,\infty}(p)= \frac{J_0}{N} \frac{(e^{\gamma \Delta_0}-1)}{(e^{\gamma \Delta_0 p}-1)}\sum_{\mu=1}^{p}\xi_i^\mu\xi_j^\mu e^{(\mu-1)\Delta_0\gamma}\ . \label{12} 
\end{eqnarray}
Eq. (\ref{J_inf}) shows explicitly that the learning protocol allows the couplings to approach in the long run a weighted version of the Hopfield couplings where each pattern's weight is a function of its presentation order ($\mu$). This means that we expect an uneven storing of the patterns: the pattern last seen is remembered best, while the memory of the previous ones decays exponentially on a time scale $\tau_{\gamma}$, in a similar way as in some generalized Hopfield models of forgetful memories \cite{nadal1986networks, parisi1986memory,  van1988forgetful}.
Finally 
expanding equation (\ref{J_inf}) for small $\gamma\Delta_0$ (many repetitions of news presented within a memory time) we obtain
\begin{equation}
\label{27}
J_{ij,\infty}= \frac{J_0}{pN} \sum_{\mu=1}^{p}\xi_i^\mu\xi_j^\mu + \frac{J_0}{pN} \sum_{\mu=1}^{p}\xi_i^\mu\xi_j^\mu (\mu-1)\Delta_0\gamma + o((\Delta_0\gamma)^2)\ .
\end{equation}
Note that the first term is proportional to the Hopfield couplings (see Eq. (\ref{10})), 
which can be thus thought as a zeroth order approximation to our couplings.\\ 
Similarly to what happens in Hopfield Neural Networks, our society will be able in the long run to store and easily retrieve the opinion patterns $\bm{\xi}^\mu$. 
The level of retrieval of the society for each of the patterns $\mu$ can be evaluated using a set of overlaps $m^\mu$: 
\begin{equation}
m^\mu(t)=\frac{1}{N}\sum_{i}\xi_i^\mu g_i(t)\ . \label{3}\\
\end{equation}
These overlaps will tell us if the system is aligned with one of the opinion configurations previously presented ($m^\mu = O(1)$) or not ($m^\mu = O(1/\sqrt{N})$). 
The value of $ m^\mu $ in the long time limit can be obtained assuming the Gaussianity of $u_i$ (as done in the scenario of the previous section)
and evaluating the average $u_i$ in the following way: we use the couplings evaluated in Eq. (\ref{12}) to evaluate the long term limit of Eq. \eqref{maineq}. In the absence of a signal ($I_i=0\  \forall i$) we thus obtain $u_i \sim \mathcal{N}(\langle u_i\rangle,\sigma_u^2)$, with
\begin{eqnarray}
\label{g_3}
\langle u_i\rangle &=& \sum_{j}J_{i,j,\infty}\langle g_j\rangle \\
\label{g_4}
\sigma_u^2 &=&\sigma^2/2\ ,
\end{eqnarray}
and 
\begin{eqnarray}
\label{g_2}
\nonumber \langle g_i\rangle &=&\erf\left(\frac{\langle u_i\rangle}{\sqrt{1+2\sigma_u^2}}\right)\\ 
 &=& \erf\left(\frac{(e^{\gamma \Delta_0}-1)}{(e^{\gamma \Delta_0 p}-1)}\frac{J_0 \sum_{\mu=1}^{p}\xi_i^\mu  m^\mu  e^{(\mu-1)\Delta_0\gamma}}
{\sqrt{1+\sigma^2}}\right).
\end{eqnarray}
If different patterns have negligible mutual overlaps, such as for uncorrelated patterns with $\frac{1}{N}\sum_{i}\xi_i^\mu \xi_i^\nu \sim \frac{1}{\sqrt{N}}$ for $\mu\neq \nu$, the equation above can have a non trivial solution for which the society aligns with exactly one of the patterns, $\nu$.
In this case $\boldsymbol{m}\simeq\{0,..0, m^\nu ,0...,0\}$ and
\begin{equation}\label{8}
 m^\nu\simeq \erf\left( m^\nu J_0 \frac{(e^{\gamma \Delta_0}-1)}{(e^{\gamma \Delta_0 p}-1)} \frac{e^{(\nu-1)\Delta_0\gamma}}{\sqrt{1+\sigma^2}}\right)\ .
\end{equation}
Note that under certain conditions the solution of Eq. (\ref{8}) is non trivial and will be larger for more recently presented patterns and smaller for older ones, meaning that, if remembered at all, recent pieces of news will be better recalled by the society.  \\
As a confirmation of this behaviour we simulated the dynamics of our society until a time $t_r$ at which we froze the couplings. 
To check whether the society has developed a memory of the $p$ external signals to which it has been exposed, after $t_r$ we apply each signal contribution again for a short time after which we remove it to observe the response of the society in terms of the overlaps $m^\mu$ in absence of it.
\begin{figure}
\centering
\includegraphics[scale=0.5]{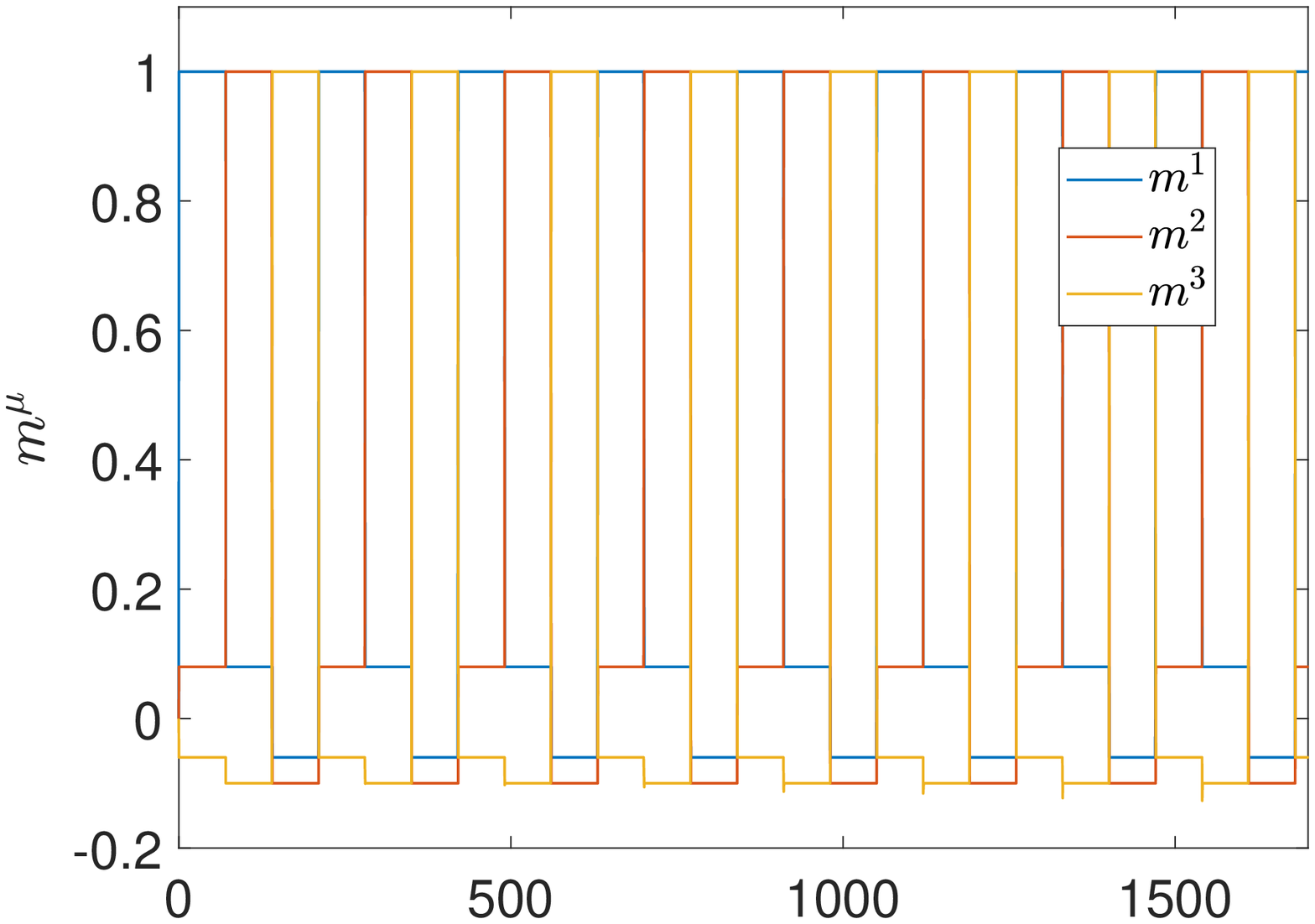}
\includegraphics[scale=0.5]{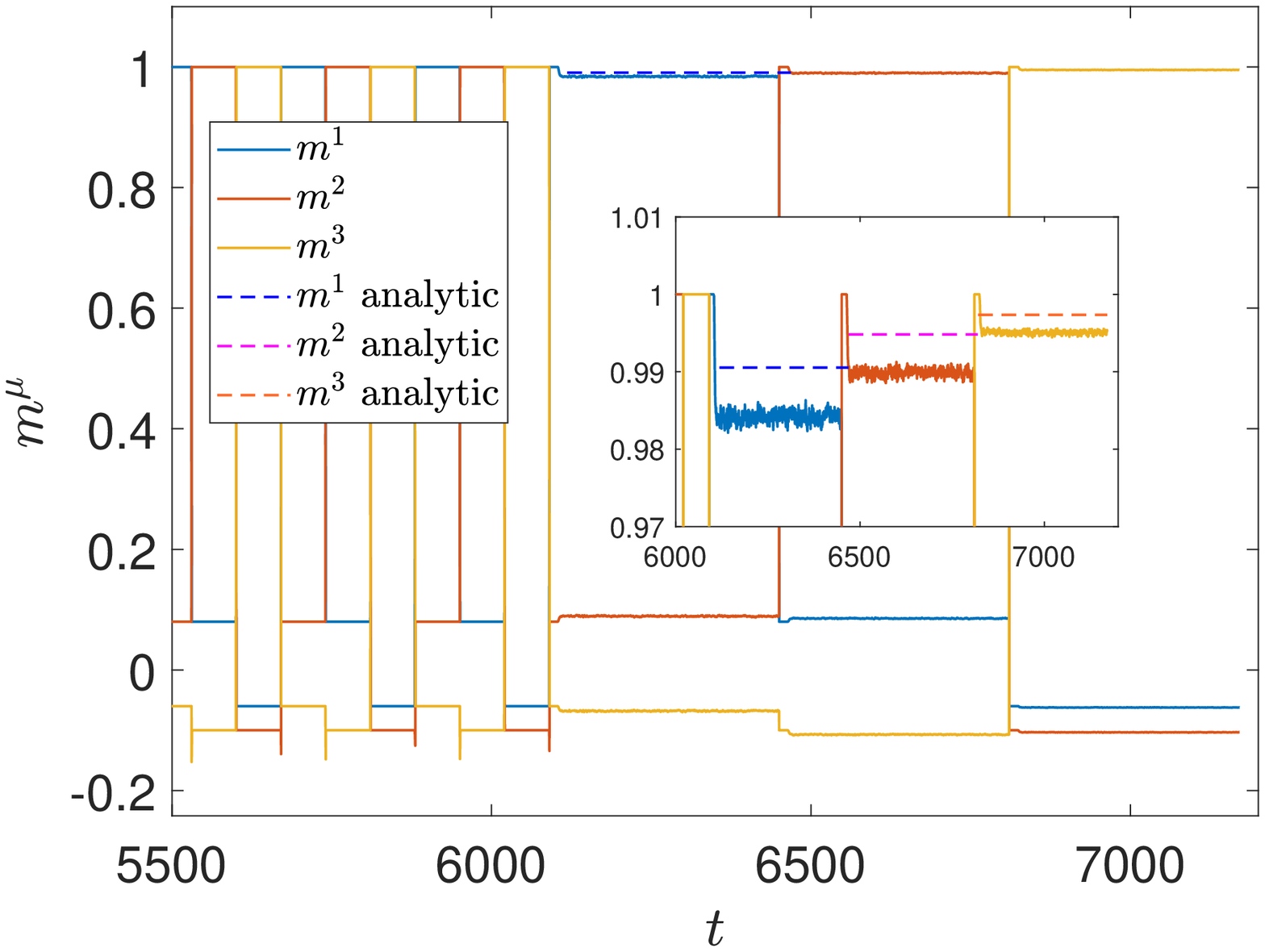}
\caption{Simulated dynamics with periodic external signal.
Upper panel: Early dynamics of the society. The overlaps with different patterns are represented by different colours and quickly reach $m^\mu\simeq 1$ when their corresponding signal contribution $\bm\xi^\mu$ is on. Lower panel: after presenting many times the patterns in a cyclic fashion, we freeze the couplings at time $t_r=6090$ and we presented each of the patterns for a time very short compared to $\Delta_0$, after which the signal is removed. The analytic predictions of the overlaps in absence of signal are compared with the simulations. In this simulation we set $I_0=50$, $J_0=6$, $\gamma=10^{-3}$ and $\Delta_0=70$. }\label{11}
\end{figure}
\begin{table}[]
\begin{tabular}{|l|lll|l|}
\hline
$\Delta_0=70$       & \multicolumn{1}{l|}{$I_0=1$} & \multicolumn{1}{l|}{$I_0=5$} & $I_0=50$                 & Analytical \\ \hline
$\overline{m^1}$ &     0.72 $\pm$ 4 $\cdot 10^{-2}$  &    0.966    
$\pm$  $ 9\cdot 10^{-3}$   &  0.97 $\pm$   1$\cdot 10^{-2}$   & 0.9905  \\ 
$\overline{m^2}$ &      0.74
     $\pm$  4 $\cdot 10^{-2}$      & 0.9894 $\pm$  4$\cdot 10^{-4}$   &  0.9887 $\pm$   $6\cdot 10^{-4}$  &   0.9948   \\
$\overline{ m^3}$ &  0.75 $\pm$ 4 $\cdot 10^{-2}$  &  0.9943 $\pm$ 3$\cdot 10^{-4}$   &   0.9947 $\pm$  2$\cdot 10^{-4}$  &  0.9973   \\ 
\hline
\end{tabular}
\caption{The table shows the values of $\overline{m^\mu}$ obtained averaging $m^\mu$ over 100 simulations with fixed $\Delta_0$ for different values of $I_0$, against their analytical predictions. The control parameter of the simulations are $J_0=6$ and $\gamma=10^{-3}$.}
\label{table}
\end{table}
As shown in Fig. \ref{11}, the society quickly reacts to each of the spikes after $t_r=6090$ as the corresponding $m^\mu$ (highlighted in Fig. \ref{11} with different colours for different signal patterns) jumps to $1$ during the spike and relaxes to a non trivial value in absence of external signal. It remains stationary until the subsequent spike of a different pattern is applied.
The expected stationary non trivial overlap $m^\mu$ obtained from Eq. (\ref{8}) matches quite well with the simulation results. To further confirm our findings we defined the quantities $\overline{m^\mu}$ as the average of $m^\mu$ over 100 simulations and we compared them for different strength $I_0$ of the signal applied during the dynamics, with their analytical predictions in Table \ref{table}. In this case to obtain $m^\mu$ in each simulation we froze the couplings at $t_r$ and we averaged the values of $m^\mu$ on the last 2000 steps after the corresponding subsequent signal spike. 
We note that predictions always slightly overestimate the simulation results. The two main reasons for this discrepancy are that in Eq. (\ref{u3}) we neglected the contributions of the fluctuations of $g_i(t)$ and that the transients of the $u_i$ dynamics after every change of external signal were neglected in the analytical evaluation of the couplings.
Moreover the theory works better for higher $I_0$ as the assumption we made that the opinions align immediately to the signal becomes more accurate for large signals.
Finally we can observe that predictions get worse for earlier external signals. Indeed they are associated with smaller effective couplings in Eq. (\ref{maineq}) and consequently overlap solutions more susceptible to the neglected fluctuations.
\begin{figure}
    \centering
    \includegraphics[scale=0.6]{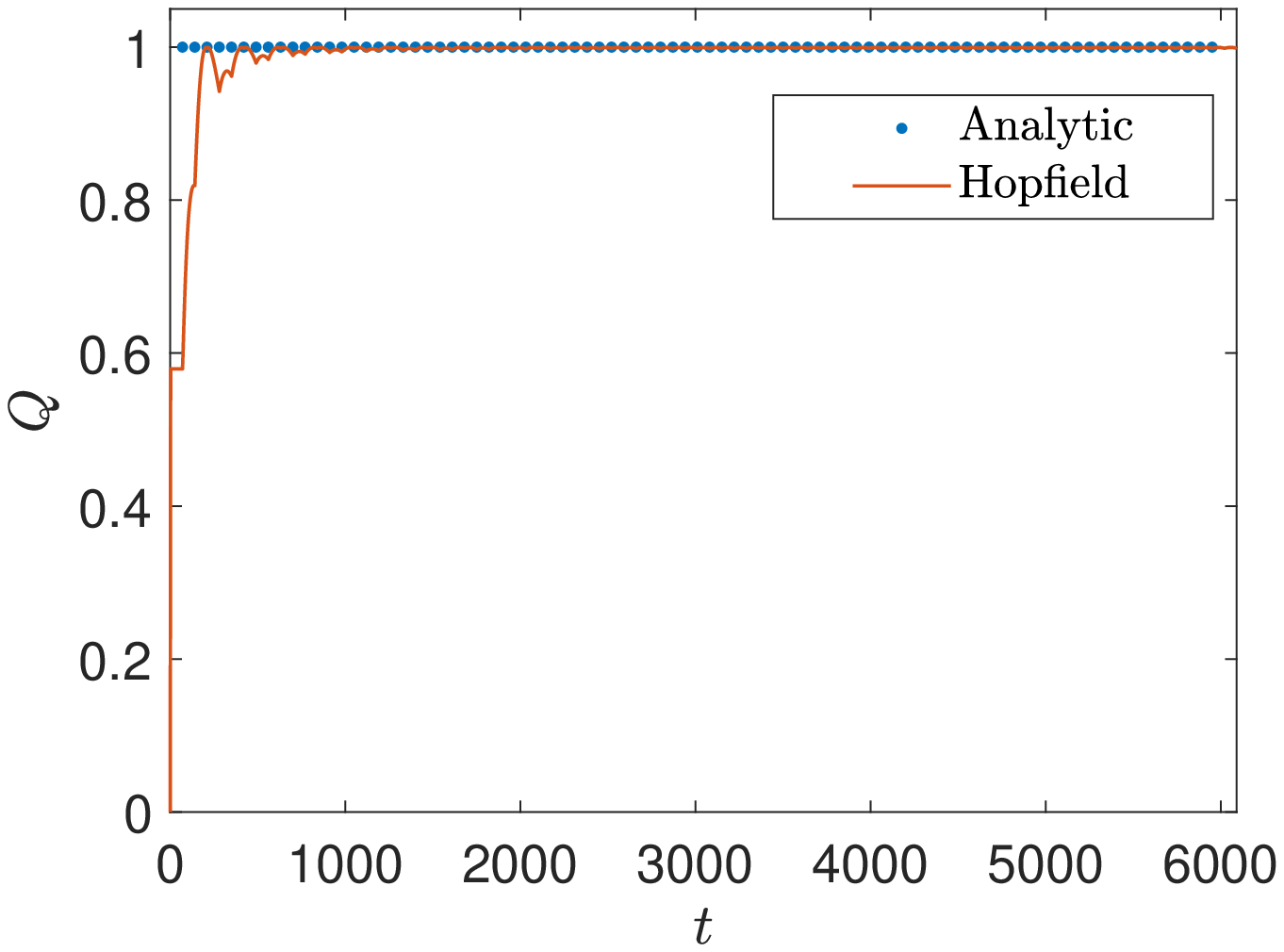}
    \includegraphics[scale=0.6]{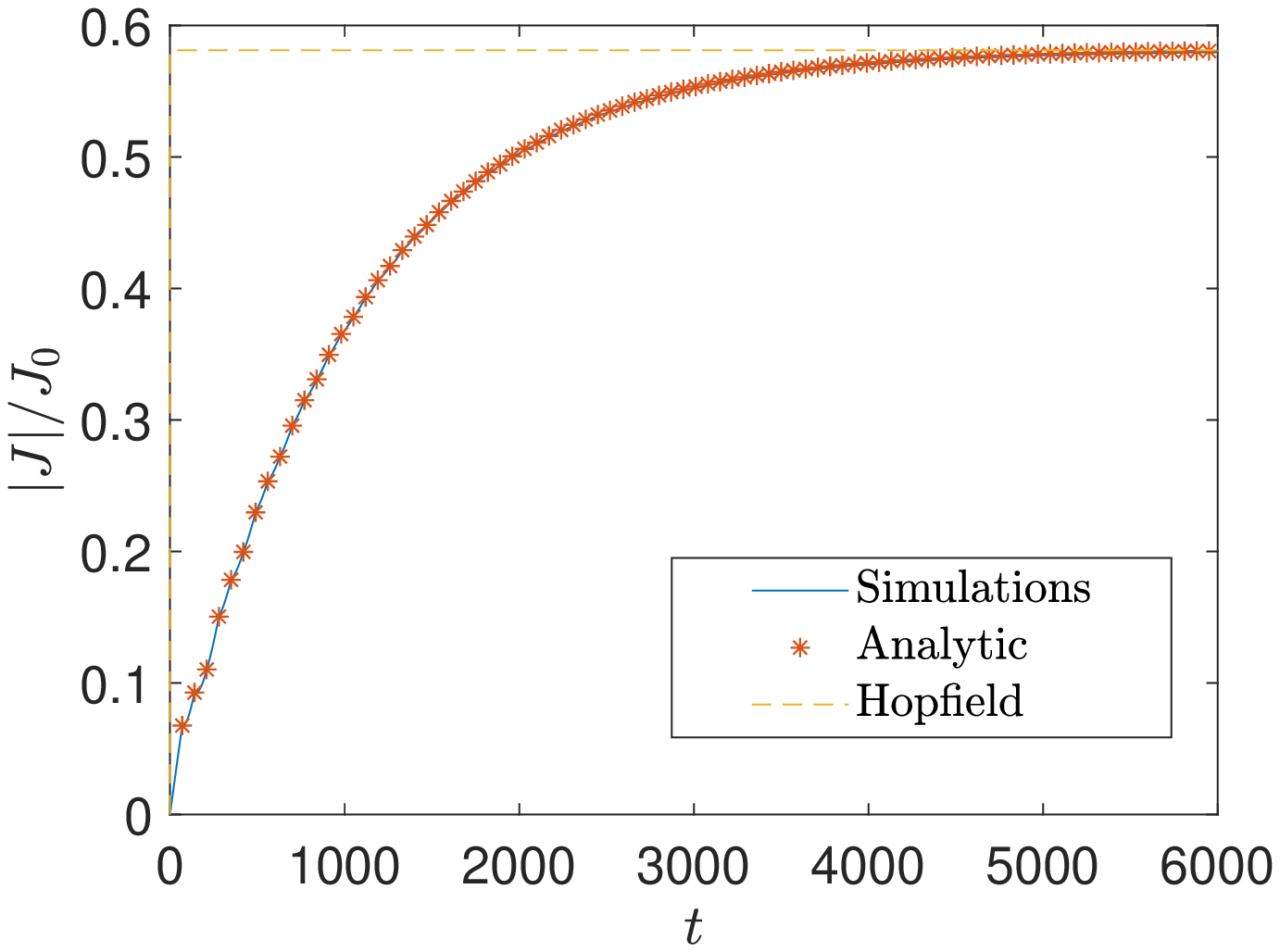}
    \caption{Simulations with periodic external signal. The upper panel shows the evolution of the correlation $Q$ between the simulated and the analytical $J$ and between the simulated $J$ and the Hopfield couplings $J^H/p$. The lower panel shows the evolution of $|J|$ against its estimated analytical evolution and the norm of the Hopfield couplings over $p$ $|J^H|/p$ in units of $J_0$.  The control parameters are the same as figure \ref{11}.}
    \label{No_drop_sig_QR}
\end{figure}\\
The possibility to store and retrieve all the presented external signals as shown in Fig. \ref{11} is expected, given the similarity between the spontaneously formed interaction couplings and classical Hopfield couplings (see Eq. \eqref{10}).
For the simulation reported in Fig. \ref{11} we indeed find that interactions very soon align almost perfectly with corresponding Hopfield couplings as shown in the upper panel of Fig. \ref{No_drop_sig_QR}. 
The figure also shows that analytic predictions of $J$ from Eq. \eqref{J_signal_nodrop_2} are very accurate as the corresponding $Q\simeq1$ at all times for $J$ analytical.
As discussed in the previous section, the possibility for the society to retrieve the pattern induced by an external signal requires sufficiently large couplings to allow a non-zero solution of Eq. \eqref{8}.
In the bottom panel of Fig. \ref{No_drop_sig_QR} we compare the norm of $J$  (defined in Eq. \eqref{absJ}) obtained with simulations to its analytical prediction and the norm of the Hopfield couplings over p $J^H/p$. The evolution of the analytical curve predicts closely the evolution of the simulated $|J|/J_0$ while $|J^H|/p$ overestimates the true value of $|J|$ at the beginning of the dynamics.
However at $t_r=6090$ the value of the simulated $|J|$ has reached a stationary value very close to $|J^H|/p$.  
The society has been irremediably shaped by the opinion patterns $\bm{\xi}^\mu$. 

\section{Intermittent external signal}
\label{sec:Learning with signal removal}
\begin{figure}
	\centering
	\includegraphics[scale=0.39]{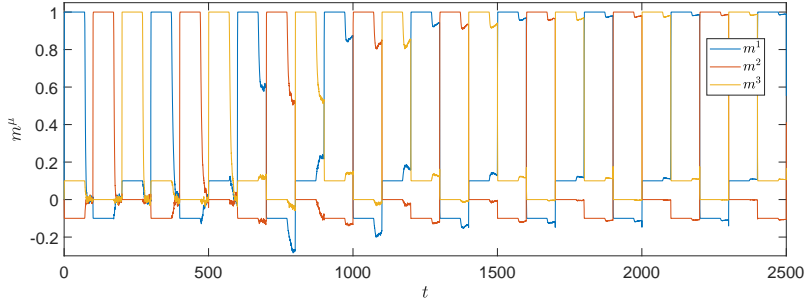}
	\caption{Simulated dynamics with intermittent signal. Three patterns are presented in a cyclic fashion. Each pattern presentation for a time $\theta \Delta_0=70$ is followed by a period $\Delta_1=30$ during which there is no external signal. Each time a pattern is presented the corresponding overlap $m^\mu$ very quickly approaches 1, and it decays to smaller values when the signal is removed. At early times, couplings are still too small to retain previously presented information and overlaps decay to small ${\cal O}(1/\sqrt{N})$ values when external signals are removed. However, after a time $t^* \simeq 600$ the couplings are able to sustain the opinion patterns even when the signal is removed. In this simulations $I_0=5, J_0=6$ and $\gamma=10^{-3}$}\label{13} 
\end{figure}

\begin{figure}
	\includegraphics[scale=0.27]{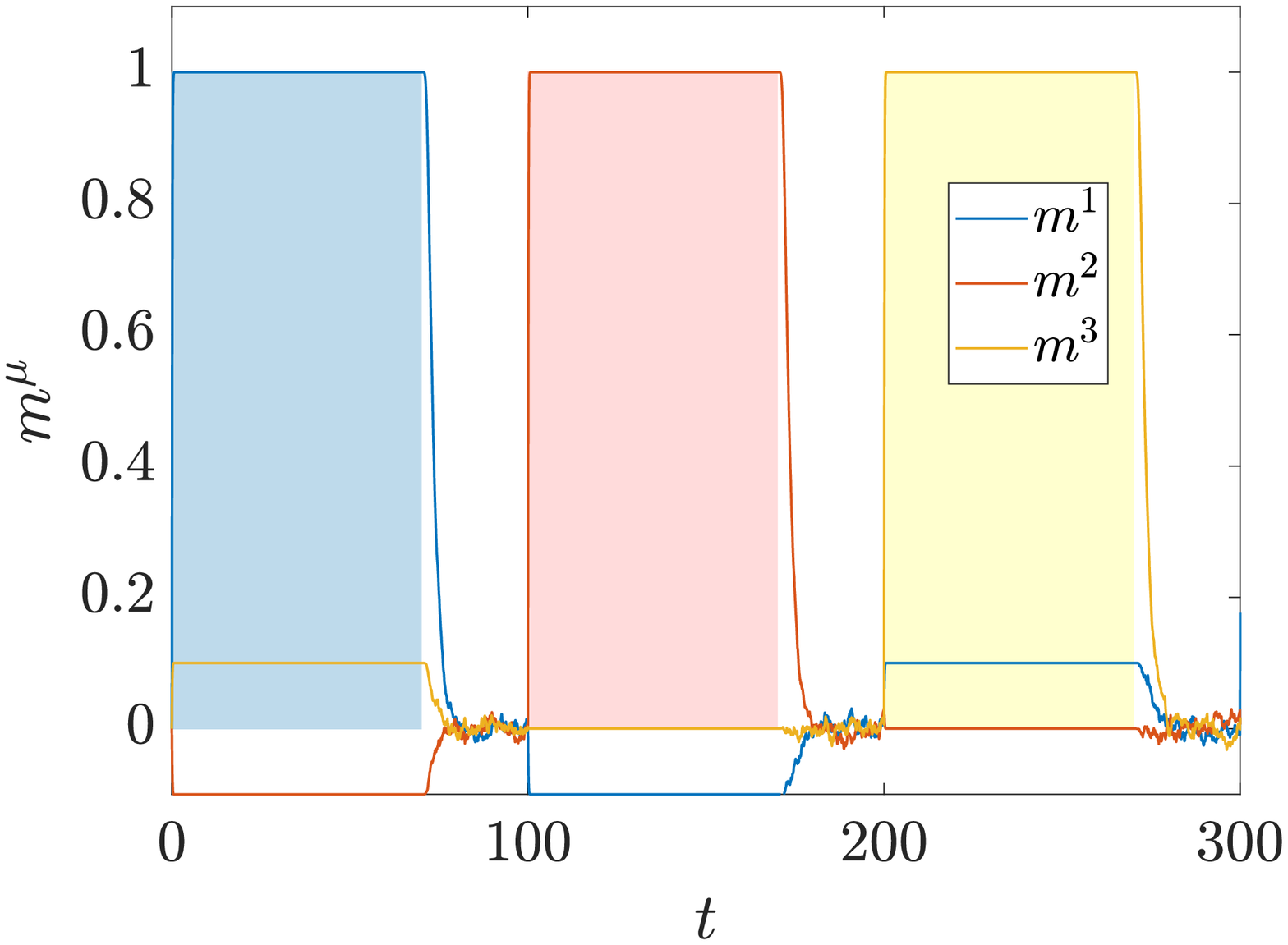}
	\includegraphics[scale=0.27]{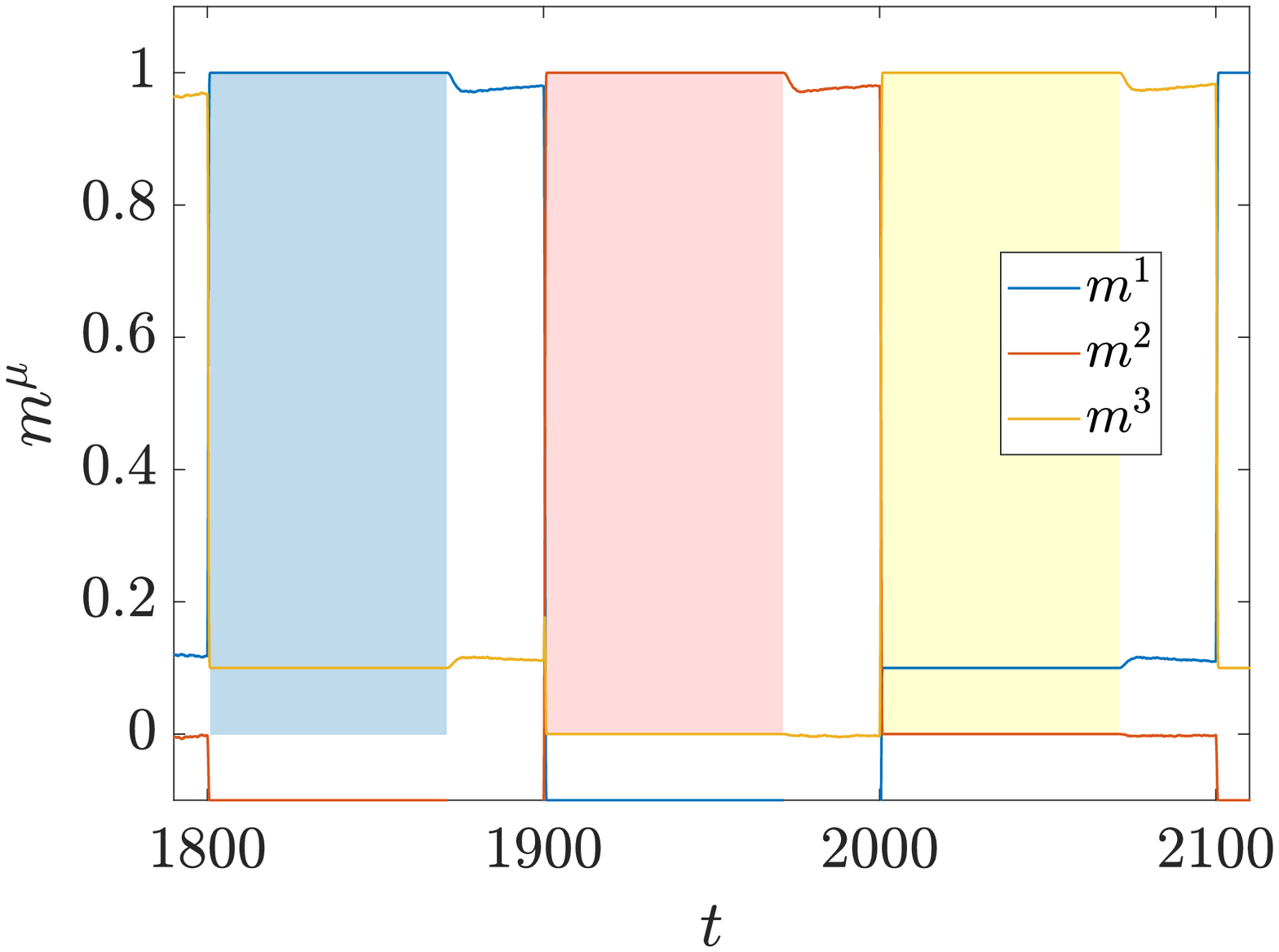}
	\caption{The figures represent a zoom into the early time regime (left panel), and into the late time regime (right panel) of Fig. \ref{13}. The coloured rectangles represent the time periods in which a signal is switched on. After the signal removal at early times the value of $m^\mu$ drops to significantly lower values, whereas it remains much close to $m^\mu=1$ at late times.} \label{33}
\end{figure}

\begin{figure}
	\centering
	\includegraphics[scale=0.6]{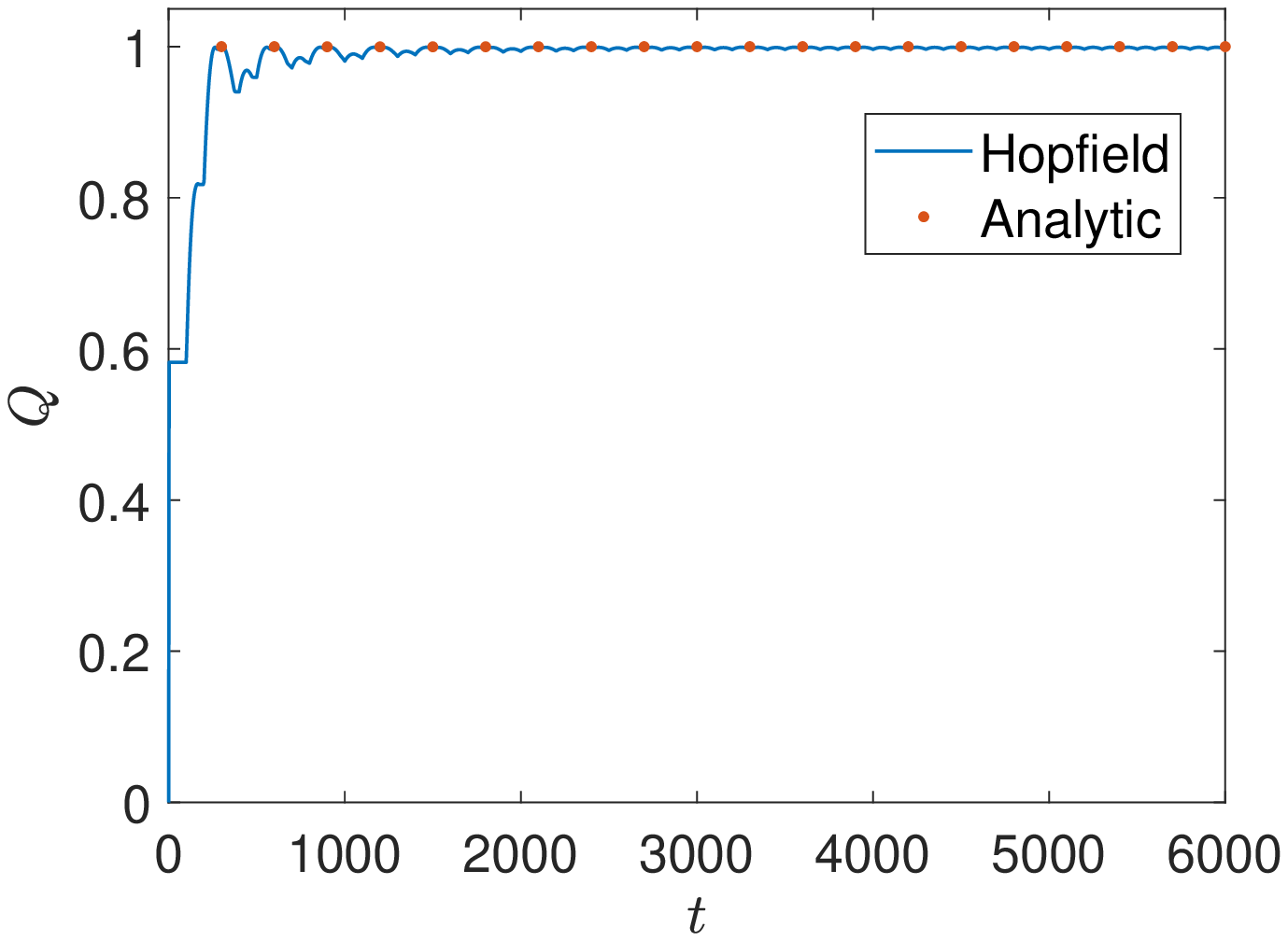}
	\includegraphics[scale=0.6]{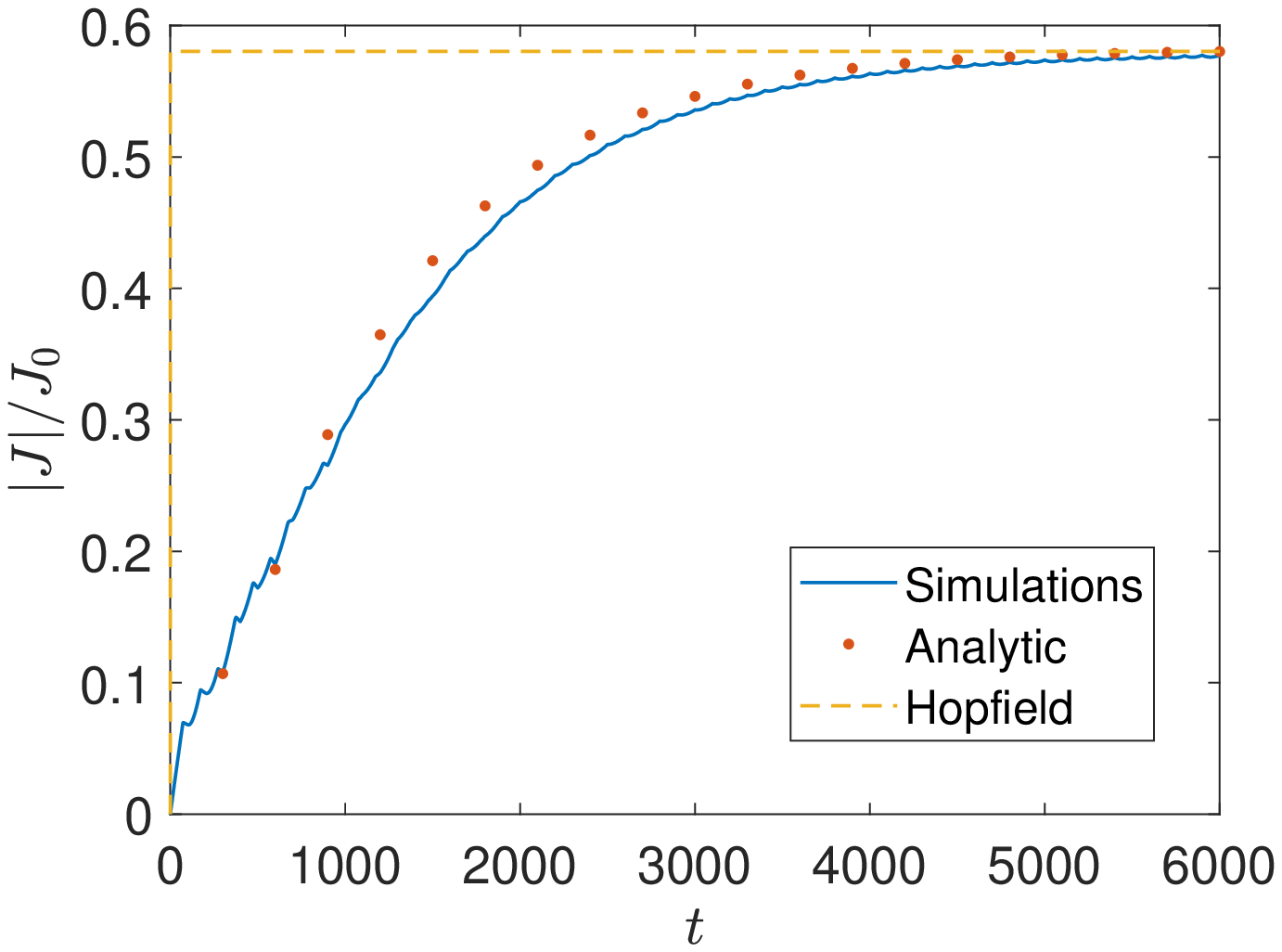}
	\caption{The upper panel shows the evolution in time of the correlation $Q$ between the simulated $J$, its analytical prediction and the Hopfield couplings over $p$, $J^H/p$. The lower panel compares the norm of $J/J_0$ from simulations to its analytical prediction and to the norm of $J^H/p$. The control parameters are the same as Fig. \ref{13}.}
	\label{sig_rem_nodrop}
\end{figure}

\begin{figure}
	\centering
	\includegraphics[scale=0.6]{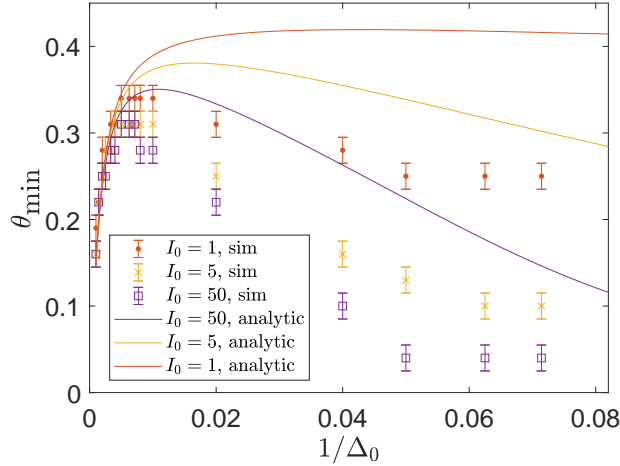}
	\caption{Simulated relation between the minimum $\theta$ necessary for the recovery of at least one pattern and the inverse of the time $\Delta_0$ compared with their analytical estimation for $\gamma=10^{-3}$ and different signal strengths.}\label{7}
\end{figure}

In this section we study a still stylized but slightly more complicated scenario. We analyse the response of the society to intermittent external information. We keep the cyclic presentation mode described in the previous section. However, the different opinion patterns are no longer influencing the society for the entire presentation period $\Delta_0$, but only for a fraction $\theta \Delta_0$ of each period, with $\theta<1$. 
The signal is absent for the remainder $\Delta_1=(1-\theta)\Delta_0$ of the presentation period:
\begin{eqnarray}\label{sig_rem}
I_i(t)= 
\begin{cases} 
I_0 \xi_i^{\mu(t)} \hspace{2cm} &n\Delta_0 < t \le n\Delta_0 + \theta \Delta_0\\
0  \hspace{2cm} &\mbox{otherwise}
\end{cases}
\end{eqnarray}
with $n \in \mathbb{N}$ and 
\begin{equation}\label{sig_rem2}
\mu(t)= 1+\left \lfloor{\frac{t}{\Delta_0}}\right \rfloor \hspace{-0.cm}\mod{p}\ .
\end{equation}
In this way we represent a society hit by a periodic sequence of different strong stimuli, such as repetitive political propaganda or a series of shocking events (e.g. terrorist attacks) alternated with periods of absence of external information. Questions that arise in such a scenario are: Will the society be shaped by these shocks? What is the smallest fraction $\theta$ of the time for which the system is exposed to external stimuli that still allows the society to spontaneously retain the information presented? \\
To evaluate the couplings in this case, we assume that during the time $\theta\Delta_0$ in which the signal $I_0{\bm \xi}^\mu$ is on, the preference fields immediately align and $\bm{g} = \bm{\xi}^\mu$, which requires the signal strength $I_0$ to be sufficiently large. As long as the society remains unable to retain the presented patterns, we find that, as soon as the signal is removed, the preference fields very quickly decay to $0$ and remain small during the time $\Delta_1$ in which the signal is off. Once the society has been exposed to sufficiently many presentation cycles, couplings of sufficient strength may have developed allowing the society to retain information about the latest pattern presented, even when the signal is removed. When evaluating the couplings for this situation, we assume for simplicity that the system remains nearly fully aligned with the previously presented pattern, $\bm{g}\simeq\bm{\xi}^\mu$, even after the the signal is turned off.
Figure \ref{13} shows a simulation exhibiting a transition from an early time regime, where information is not retained after signal removal, to a late time regime, where the system remains aligned with a signal even at times where the signal is switched off. In Fig. \ref{33}, we present a zoom into both the early time and the late time regimes. Shaded rectangles in the figure represent the intervals $\theta\Delta_0$ in which an external signal is present. At early time when the signal is switched off, the opinions take some time to disalign to it. This extra time, that we will indicate as $\theta' \Delta_0$, is not easy to calculate, however we can give a rough estimation of it assuming that the preference field $u_i(t)$ freely decays to zero when the signal is removed (see \ref{appendix_thetaprime}).
We define the time $t^*$, in multiples of $p\Delta_0$, as the time of the last cycle of external stimuli that is still insufficient to create couplings of a strength needed for the society to retain the previously presented information.
Using the signal structure defined in Eq. \eqref{sig_rem}, we can derive an expression for the couplings for $t<t^*$ and for $t>t^*$ under the simplifying assumptions made above (details of the calculations in \ref{appendix_drop}):
\begin{eqnarray}\label{J_a_1}
J_{ij}(t<t^*)=\frac{J_0}{N}(\e^{\gamma\Delta_0(\theta+\theta')}-1)\e^{-\gamma t}\frac{(1-e^{\gamma t})}{(1-e^{\gamma \Delta_0 p})} \sum_{\mu=1}^p \xi_i^\mu \xi_j^\mu \e^{\Delta_0\gamma(\mu-1)}
\end{eqnarray}
\begin{eqnarray}\label{J_a_2}
J_{ij} ( t>t^*) &=&\nonumber \frac{J_0}{N} e^{-\gamma t}\left((e^{\gamma \Delta_0(\theta+\theta')}-e^{\gamma \Delta_0})\frac{(1-e^{\gamma t^*})}{(1-e^{\gamma \Delta_0 p})} \right. \nonumber \\
&\ & +  \left. (e^{\gamma \Delta_0}-1)\frac{(1-e^{\gamma  t})}{(1-e^{\gamma \Delta_0 p})}\right) \sum_{\mu=1}^{p}\xi_i^\mu\xi_j^\mu e^{(\mu-1)\Delta_0\gamma}   \nonumber \\ 
\end{eqnarray}
The couplings in Eq.s \eqref{J_a_1} and \eqref{J_a_2} are for simplicity evaluated only for integer multiples of $\Delta_0$.
As shown in Fig.  \ref{sig_rem_nodrop}, the couplings thus predicted compare remarkably well with those evaluated in a numerical simulation of the dynamics as presented in Fig. \ref{33}. However, given the approximations used in the estimation of $\theta'$ we cannot expect to have a perfect agreement between the analytical prediction and the simulations (the limitations of our approach are discussed in \ref{appendix_I0}).
In Fig. \ref{sig_rem_nodrop},  the choice of parameters is such that the interactions rapidly align with the Hopfield couplings, and their norm grows along the dynamics eventually granting retrieval of the signal patterns after a finite time $t^*$. \\
To determine $t^*$, we use these equations to obtain an expression for the couplings $J$ at time $t^*+(\theta+\theta') \Delta_0$ (see \ref{appendix_drop} for details of the calculation) and use these in the self-consistency equation for $m^1$ (also derived in the same appendix)
\begin{equation}
\label{m_sig_drop}
 m^1 =\mbox{erf} \left( m^1 \frac{J_0(1-\e^{-\gamma\Delta_0(\theta+\theta')})}{\sqrt{(1+\sigma^2)}} \frac{(1-e^{-\gamma (t^*+\Delta_0 p)})}{(1- e^{-\gamma \Delta_0 p})}\right)\ .
\end{equation}
The value of $t^*$ is then obtained by requiring that Eq. \eqref{m_sig_drop} has a non trivial solution, which gives
\begin{eqnarray}
\label{t*}
t^*=  -\Delta_0p \left[1 + \frac{1}{\gamma \Delta_0 p }\log \left(1 - \frac{\sqrt{(1+\sigma^2)}}{J_0(1-\e^{-\gamma\Delta_0(\theta+\theta')})}\frac{\sqrt{\pi}}{2}(1-e^{-\gamma \Delta_0 p})   \right) \right] \ .
\end{eqnarray}
Note that $t^*$ increases with decreasing $\theta$, and it will eventually diverge (and a finite $t^*$ will cease to exist)
as $\theta$ is decreased below $ \theta_{\mbox{min}}$
\begin{eqnarray}
    \theta_{\mbox{min}}&=&-\frac{1}{\gamma \Delta_0}\log\left(1-\frac{\sqrt{\pi(1+\sigma^2)}}{2J_0}(1-\e^{-\gamma\Delta_0 p})\right)-\theta'\\ \nonumber
    &=& -\frac{1}{\gamma \Delta_0}\log\left(1-\frac{\sqrt{\pi(1+\sigma^2)}}{2J_0}(1-\e^{-\gamma\Delta_0 p})\right)- \frac{1}{\Delta_0}\log \left(\frac{I_0(1-e^{-\Delta_0\theta})}{0.74}\right)
\end{eqnarray}
for which the argument of the logarithm in Eq. \eqref{t*} vanishes and where $\theta'$ is evaluated in \ref{appendix_thetaprime}.
 The solution of this equation can be found numerically and the resulting behaviour in function of $1/\Delta_0$ is shown in Fig. \ref{7}. The condition $\theta > \theta_{\mbox{min}}$ thus guarantees the existence of a finite time $t^*$ at which persistent memory starts to form, and at least one of the patterns stored can be recovered\footnote{For $\theta > \theta_{\mbox{min}}$ at least one pattern will be recovered, but this does not guarantee that the first pattern of the cyclically repeated sequence is among those recovered.}. 
The existence of a minimum value of $\theta$ required for the society to be able to spontaneously retrieve the information contained in the signals presented earlier is of immediate practical relevance. For advertisement campaigns, for instance, it defines the minimum fraction of time needed for a repeatedly presented signal to permanently impress the audience as a collective body. In the domain of news, it would, for instance allow to assess, whether or not repeated news items might leave a subtle persistent trace in the society and produce collective responses otherwise unpredictable.\\
To verify our predictions of $\theta_{\mbox{min}}$ we simulated dynamics with external signal of different amplitude $I_0$ until stationary interaction couplings are reached. We then froze the couplings and counted how many times the society recovers at least one of the patterns\footnote{This includes also mixture states with $\mathbf{m}=(m^1, m^2, m^3)$, where $m^1\ne 0$, $m^2\ne 0$ and $m^3\ne 0$, which are also encountered in some instances.} after a signal spike. 
Recovery is reached when the corresponding overlap in absence of external signal satisfies the threshold condition $m^\mu>0.4$.
Such recovery threshold has been chosen significantly higher than the overlap ($\sim 1/\sqrt{N} = 0.1$) expected if the system state is uncorrelated with the pattern, but not too high in order not to exclude recovery with a low $O(1)$ overlap given the system parameters. We finally estimate $\theta_{\mbox{min}}$ from simulations as the smallest $\theta$ for which at least half of $50$ trial runs of the dynamics show such retrieval behaviour.
In Fig. \ref{7} we plot $\theta_{\mbox{min}}$ for different values of $I_0$ as a function of the inverse total presentation time $1/\Delta_0$ alongside the analytic prediction $\theta_{\mbox{min}}$ obtained above.  \\
As clearly visible in the figure, for large $\Delta_0$ the analytic curve gives a good prediction of the simulation results for all the signal strengths, confirming that there is not significant dependence on $I_0$ in this regime. It is interesting to notice that in this regime the value of $(\theta_{\mbox{min}}+\theta')\Delta_0$ corresponds to the minimum time needed by the society to embed a single pattern presented with a persistent signal (the minimum $t_r$ mentioned in section \ref{sec:per_ext_sig}):
\begin{equation}
    \theta_{\mbox{min}}+\theta'\simeq -\frac{1}{\gamma \Delta_0}\log\left(1-\frac{\sqrt{\pi(1+\sigma^2)}}{2J_0}\right)\ 
\end{equation}
assuming $J_0>\sqrt{\pi(1+\sigma^2)}/2$. This means that at the beginning of the dynamics the society is able to maintain its orientation towards the very first pattern seen, that will be the one most easily remembered. \\
Interestingly at small $\Delta_0$ a much stronger dependence on $I_0$ develops. The analytical curves qualitatively capture the trend of the numerical ones but overestimate their true value. This discrepancy can be related to the approximations used in the estimate of $\theta'$. 
When solving the dynamics in the fraction of time $\theta'\Delta_0$ we neglect the interactions between agents.
In doing this we underestimate the partial memory that has started to form in the society, and consequently overestimate the $\theta_{\mbox{min}}$ needed for recovery. \\
Lastly but very importantly we note that both numerical results and analytic estimations show a decrease in $\theta_{\mbox{min}}$ at small $\Delta_0$ which is more pronounced when $I_0$ is larger.
This phenomenon is a direct consequence of the fact that $\theta'$ increases with $I_0$, therefore for larger $I_0$ the term $\Delta_0\theta'$ allows recovery with smaller $\theta_{\mbox{min}}$. 
In applications the fact that 
$\theta_{\mbox{min}}$ reaches very small values for small $\Delta_0$ and large $I_0$ would suggest to invest in short but frequent and high-impact advertisements. Similarly we can conclude that shocking events such as terror attacks, if repeated on short periods, might leave deep traces in the society despite being very localised in time.

 \section{Conclusions}
 News of disruptive events in history such as terror attacks often appear to change the behaviour of a society and will influence how people will react to news of future events of a similar kind. In this work we introduce a simple model of opinion dynamics which includes otherwise well-studied phenomena such as homophily \cite{lazarsfeld1954friendship, mcpherson2001birds, axelrod1997dissemination, deffuant2000mixing, hegselmann2002opinion} (the tendency of people to interact more often with others who share similar opinions) and xenophobia \cite{macy2003polarization, flache2011small, baldassarri2007dynamics,  mark2003culture} (the tendency to adopt  opinions different different from those of people with whom there has been disagreement in the past). The model includes dynamically evolving couplings, which effectively record an exponentially weighted history of co-expressed opinions between any pair of agents in the system. We show how this mechanism allows a society do develop a collective memory of external information it had previously been exposed to, allowing it to spontaneously retrieve such information in the future when briefly triggered by exposure to that information.\\ 
  We study the emergence of this type of collective memory both analytically and by way of simulations for three stylized scenarios representing different histories of exposure to external information: (i) information consisting of a persistent signal, (ii) information consisting of repeated presentations of a set of different signal patterns, and (iii) information consisting of repeated presentations of a set of different signal patterns separated by periods of absence of any signal. In the first scenario, the external information does not change in time; the society aligns to the signal and --- after a sufficiently long time of exposure to the signal --- will remember it in the future even when the signal is removed. In the second scenario, the society is exposed to a series of signals, corresponding to different news. If these news are repeated for a sufficient number of times, the society is able to remember all of them, and recall them in the future when triggered by a brief spike of the same information. This can be true also in the third scenario, in which the different news are are interspersed with phases of absence of any signal. The determining factor here is the relative length of the periods of signal presentation and signal removal. We were able to compute the critical minimal ratio of presentation time and signal removal time that allows the society to develop a persistent memory of the sequence of news and thereby to remain aligned to external information even when the signal is removed. Moreover we demonstrated that even very short signals, if sufficiently strong and repeated sufficiently often, can guarantee the spontaneous retrieval of their information.\\
 In the three scenarios analysed, polarizing signals presented to the society were able to deeply change the collective behaviour of the society described by our model, in the sense that persistent memory of past events emerged which causes it to react differently in the future. The condition for this to occur is always that the bare coupling constant $J_0$ in the model is sufficiently large compared to the noise level of the dynamics. Thus our model is able to capture, how the collective behaviour of a society can be strongly influenced by its past events.\\
 A follow up work will concern the study of the model under more realistic assumptions about signal structures, including presentation of news items in random order and presentation of news with different signal intensities. An interesting further generalization that could be considered to make the model more realistic is to define interactions depending on  multidimensional (rather than binary scalar) opinions so as to represent the effect that individuals interact in ways which depend on judgments about a collection of topics. In this setting the evolving interactions based on past interpersonal agreement or disagreement on the entire set of topics would correlate the agents' response to the different topic in a non trivial way.
 
\section*{Acknowledgements}
The authors acknowledge funding by the Engineering and Physical Sciences Research Council  (EPSRC)  through  the  Centre  for  Doctoral  Training  in  Cross  Disciplinary Approaches to Non-Equilibrium Systems (CANES, Grant Nr.  EP/L015854/1). The authors would like to thank Nishanth Sastry for his contribution at the initial stage of this project and Jean-Philippe Bouchaud, Ton Coolen, Imre Kondor and Francesca Tria for interesting discussions.

\appendix
\section{Analytical prediction of $J_{ij}$ for a repetitive signal}\label{appendix_nodrop}
We want to give an analytical estimation of $J_{ij}$ for a signal defined in Eqs. (\ref{18}) and (\ref{32}). For simplicity, we will evaluate the couplings only at the end of each pattern presentation period (i.e., for times $t$ which are integer multiples of $\Delta_0$). We thus define $J_{ij}(\nu,t)$ as the coupling at time $t=N_p p\Delta_0+\nu \Delta_0$, in which $N_p$ is the number of complete cycles of $p$ patterns and $\nu < p$ the number of additional patterns seen in the final (possibly incomplete) cycle presented up to time $t$. If we assume that as soon as the signal is switched on, the expressed opinions $\bm{g}$ will for each $\mu$ presented align with the pattern $\bm{\xi}^\mu$,  then $J_{ij}(\nu, t)$ will take the form
\begin{eqnarray} \label{J_nodrop_1}
J_{ij}(\nu, t) &=&\frac{J_0}{N}\gamma\int_{0}^{t}\D s e^{-\gamma(t-s)}g_i(s)g_j(s)\label{19}\nonumber\\ \nonumber
&=& \frac{J_0}{N}\gamma \left[\left( \xi_i^1\xi_j^1 \int_{0}^{\Delta_0}\D s e^{-\gamma(t-s)}+...\right.\right.\\
\nonumber &\ & \left.+ \xi_i^p\xi_j^p \int_{(p-1)\Delta_0}^{p\Delta_0}\D s e^{-\gamma(t-s)}\right) \\ \nonumber
&\ &+  \left( \xi_i^1\xi_j^1 \int_{p\Delta_0}^{(p+1)\Delta_0}\D s e^{-\gamma(t-s)} +...\right.\\ 
&\ & +\left. \xi_i^p\xi_j^p \int_{(2p-1)\Delta_0}^{2p\Delta_0}\D s e^{-\gamma(t-s)}\right) + . . .\nonumber \\  \nonumber
&\ & + \left. \left( \xi_i^1 \xi_j^1 \int_{N_p p\Delta_0}^{(N_p p+1)\Delta_0}\D s e^{-\gamma(t-s)} +...\right.\right.\\ 
&\ &+ \left. \left. \xi_i^\nu \xi_j^\nu \int_{(N_pp+\nu-1)\Delta_0}^{t=(N_pp+\nu)\Delta_0}\D s e^{-\gamma(t-s)}
\right) \right]\ .
\end{eqnarray}
Using
\begin{equation}\label{J_nodrop_2}
\int_{a}^{a+\Delta_0} e^{\gamma s}\D s=\int_{0}^{\Delta_0}e^{\gamma x+\gamma a}dx=\frac{(e^{\gamma \Delta_0}-1)e^{\gamma a}}{\gamma}\ ,
\end{equation}
we obtain
\begin{eqnarray}\label{J_nodrop_3}
J_{ij} &=& \frac{J_0}{N} e^{-\gamma t} (e^{\gamma \Delta_0}-1)
\left[\left( \xi_i^1\xi_j^1 +. . .+\xi_i^p\xi_j^p  e^{\gamma (p-1)\Delta_0}\right)\right.
\nonumber
\\ \nonumber
&+&  \left. \left( \xi_i^1\xi_j^1 e^{\gamma p\Delta_0}+. . .+\xi_i^p\xi_j^p  e^{\gamma (2p-1)\Delta_0}\right)+ . . .\right.\nonumber \\
&+&  \left. \left( \xi_i^1\xi_j^1 e^{\gamma N_p p\Delta_0}+. . .+\xi_i^\nu\xi_j^\nu  e^{\gamma (N_pp+\nu -1)\Delta_0}\right)\right]\ .
\end{eqnarray}
Now we can group the terms and using $t=N_pp\Delta_0 + \nu\Delta_0$ we have
\begin{eqnarray} \label{J_nodrop_4}
J_{ij}(\nu,t) &=&  \frac{J_0}{N} (e^{\gamma \Delta_0}-1)\left(e^{-\gamma t}  \sum_{\mu=1}^{p}\xi_i^\mu\xi_j^\mu e^{(\mu-1)\Delta_0\gamma}\sum_{k=0}^{(t-\nu\Delta_0)/(p\Delta_0)-1} e^{ \gamma \Delta_0 pk}\right. \nonumber\\
&\ & + \left. \sum_{\mu=1}^\nu \xi_i^\mu\xi_j^\mu e^{-(\nu-\mu+1)\Delta_0\gamma}\right) \nonumber \\
&=&\nonumber \frac{J_0}{N}  (e^{\gamma \Delta_0}-1)\left( \frac{(e^{-\gamma t}-e^{ -\gamma\nu\Delta_0)})}{(1-e^{\gamma \Delta_0 p})}\sum_{\mu=1}^{p}\xi_i^\mu\xi_j^\mu e^{(\mu-1)\Delta_0\gamma} \right.\\
&\ &+ \left. \sum_{\mu=1}^\nu \xi_i^\mu\xi_j^\mu e^{-(\nu-\mu+1)\Delta_0\gamma}\right)\ .
\end{eqnarray}

\section{$\theta'$ estimation for intermittent signal} \label{appendix_thetaprime}
We would like to have an estimate of the the time needed by the expressed opinions to disalign with the signal when this is removed. Given that we are not able to evaluate the integral in Eq. \eqref{eqJs} during the decaying transient of $g_i(t)$, we can estimate it considering $|g_i(s)g_j(s)|=1$ for a time $(\theta+\theta')\Delta_0$ and 0 for the remaining time $\Delta_0(1-\theta-\theta')$. The time $\theta' \Delta_0$ can be estimated as the time in which $|g_i(s)g_j(s)|$ falls to half of its value at the removal of the signal (that is approximately 1). In this way the overestimation implied by assuming $|g_i(s)g_j(s)|=1$ for $s \le \theta'\Delta_0 $ is compensated by the underestimation made by assuming $|g_i(s)g_j(s)|=0$ at subsequent times. We can give a rough estimate of $g_i$ during the decay assuming that the signal is switched on at time 0 and as soon as it is removed the preference fields follow:
\begin{equation}\label{ui}
    u(t) = \langle u(\Delta_0\theta)\rangle \e^{-t}\ .
\end{equation}
where we recall that  $u = \xi_i u_i$ and $u_i(\theta\Delta_0)$ is the solution of Eq. \eqref{maineq} just before the signal is removed. Here we also assume that the interactions between the agents and the noise are neglected, such that the preference field freely decay to 0.  Using this, we can calculate the time $t=\theta'\Delta_0$ for which $|g_i(t)g_j(t)|=0.5$ that is:
\begin{equation}
\label{theta_prime}
    \theta' \simeq \frac{1}{\Delta_0}\log ( \langle u(\Delta_0\theta)\rangle /0.74)\ .
\end{equation}
In order to find $u_i(\theta\Delta_0)$ we use Eq. \eqref{u3} and we assume that the opinions are 0 before the signal is presented and they align to it as soon as it is switched on with $g_i \sim \xi_i^\mu$ . As a consequence in Eq. \eqref{u3} we will have $u_i(0)=0$ and
\begin{eqnarray}
   \langle U _i(\Delta_0\theta)\rangle &=&  \frac{\gamma}{N} \int_0^{\Delta_0\theta} \mbox{d} s \ \e^{-\gamma(\Delta_0\theta-s)}g_i(s) \sum_j g_j(s)g_j(\Delta_0\theta) \nonumber\\
   &=&  \xi_i^1  m(t)(1-e^{-\gamma \Delta_0\theta})\ ,
\end{eqnarray}
which in the limit $\Delta_0 \ll \tau_\gamma$ which we consider in this paper gives:
\begin{equation}\label{<U>}
    |\langle U _i(\Delta_0\theta)\rangle| = |m(t)\gamma \Delta_0\theta| \ll 1
\end{equation}
and so it is negligible respect to the other terms in the equation.
 The final estimation of $\langle u(\Delta_0\theta)\rangle$ will thus be:
 \begin{equation}
    \langle u(\Delta_0\theta)\rangle = I_0(1-e^{-\Delta_0\theta})\ ,
 \end{equation}
 that inserted in Eq. \eqref{theta_prime} gives:
\begin{eqnarray}
    \theta'&=& \frac{1}{\Delta_0}\log \left(\frac{I_0(1-e^{-\Delta_0\theta})}{ 0.74}\right)\ .
\end{eqnarray}
We want also to remark that neglecting the terms in Eq. \eqref{<U>} from the calculations does not influence significantly the final result. In fact if we substitute $m \simeq 1$ in Eq. \eqref{<U>} and we use this to estimate $\theta'$ the results that we obtain once the other parameters inserted the results do not differ substantially when compared to those obtained by neglecting this subdominant contribution. 

\section{$J_{ij}$ for intermittent signal} \label{appendix_drop}
Given a signal of the form in Eq. (\ref{sig_rem}) and \eqref{sig_rem2} we want to calculate the couplings $J_{ij}$ in our model.  In order to do this we will assume that at the beginning of the dynamics, as soon as soon as each signal contribution $\mu$ is switched on, the preference field aligns to it with $g_i = \xi_i^\mu$. The opinions will remain aligned to the signal for a time $\Delta_0(\theta + \theta')$ where $\theta\Delta_0$ is the time the signal is actually on and $\theta' \Delta_0$ is the additional time the opinions remain aligned to the signal during the decay of the preference field $g_i$.\\
Let us define $t=N_p\Delta_0 p$, with $N_p$ as the total number of complete cycles of $p$ pattern presentations seen at time $t$. 
Following the same reasoning of \ref{appendix_nodrop} we will calculate $J$ at time $t$ as: 
\begin{eqnarray}
J_{ij} &=& J_0 \frac{\gamma}{N} \int_0^{t}\e^{-\gamma(t-s)}g_i(s)g_j(s) \nonumber\\
&=&  J_0 \frac{\gamma}{N} \left( \xi_i^1 \xi_j^1 \int_{0}^{\Delta_0 (\theta+\theta')} \e^{-\gamma(t-s)} ds + ... \right. \nonumber \\ \nonumber
&\ & + \left. \xi_i^p \xi_j^p \int_{\Delta_0(p-1)}^{\Delta_0 (p-1+\theta+\theta')} \e^{-\gamma(t-s)}ds\right) + ...\\ \nonumber
&\ & + \left( \xi_i^1 \xi_j^1 \int_{(N_p-1) p\Delta_0}^{\Delta_0((N_p-1) p+\theta+\theta')} \e^{-\gamma(t-s)}ds + \right. ... \\
&\ & \left. + \xi_i^p \xi_j^p \int_{\Delta_0N_p(p-1)}^{\Delta_0(N_p(p-1)+\theta+\theta')} \e^{-\gamma(t-s)}ds \right)  \label{JA_nodrop} 
\end{eqnarray}
Now using  
\begin{equation}
\label{integral}
\int_{a}^{a+\Delta_0(\theta+\theta')}\e^{\gamma s}ds= \left(\e^{\gamma\Delta_0 (\theta+\theta')}-1\right)\frac{e^{\gamma a}}{\gamma}\ ,
\end{equation}
we get
\begin{eqnarray}
J_{ij}&=& \frac{J_0}{N}\left (\e^{\gamma\Delta_0(\theta+\theta')}-1\right) e^{-\gamma t} \sum_{\mu=1}^p \xi_i^\mu \xi_j^\mu \e^{\Delta_0\gamma(\mu-1)}\sum_{k=0}^{N_p-1}\e^{\Delta_0\gamma p k} %+ \right. 
\nonumber\\
\end{eqnarray}
and finally exploiting
\begin{equation}
\sum_{k=0}^{N_p-1}\e^{\Delta_0\gamma p k}=\frac{(1-e^{\gamma\Delta_0p N_p})}{(1-e^{\gamma \Delta_0 p})}=\frac{(1-e^{\gamma t})}{(1-e^{\gamma \Delta_0 p})}
\end{equation}
we obtain
\begin{eqnarray}
J_{ij}&=&\frac{J_0}{N} (\e^{\gamma\Delta_0(\theta+\theta')}-1)\frac{(e^{-\gamma t}-1)}{(1-e^{\gamma \Delta_0 p})}\sum_{\mu=1}^p \xi_i^\mu \xi_j^\mu \e^{\Delta_0\gamma(\mu-1)} 
\ .
\end{eqnarray}

If the amplitude $J_0$ of the couplings is too small for the given noise level of the dynamics, the society may never be able to retrieve any of the information it was previously exposed to, in which case the above expression holds for all $t$. Otherwise, if $J_0$ is large enough, for some $\theta$ the society is able to retrieve the pattern 1 at time $t=t^*+\Delta_0(\theta+\theta')$ after its presentation, with $t^*=N_p^*\Delta_0 p$. For this to happen we thus need a non-trivial solution of:
\begin{equation}
m^1=\mbox{erf} \left( m^1 \frac{J_0(1-\e^{-\gamma\Delta_0(\theta+\theta')})}{\sqrt{(1+\sigma^2)}} \frac{(e^{-\gamma (t^*+\Delta_0 p)}-1)}{(e^{-\gamma \Delta_0 p}-1)} \right) \ ,
\label{m1eq}
\end{equation}
obtained using $J$ at $t=t^*+(\theta+\theta')\Delta_0$:
\begin{eqnarray}
J_{ij}&=&\frac{J_0}{N} (1-\e^{-\gamma\Delta_0(\theta+\theta')})\bigg [\frac{(e^{-\gamma t^*}-1)}{(1-e^{\gamma \Delta_0 p})}\sum_{\mu=1}^p \xi_i^\mu \xi_j^\mu \e^{\Delta_0\gamma(\mu-1)} + \nonumber\\
&\ &  + \xi_i^1\xi_j^1 \bigg ]
\ .
\end{eqnarray}
We thus need
\begin{equation}
\frac{J_0(1-\e^{-\gamma\Delta_0(\theta+\theta')})}{\sqrt{(1+\sigma^2)}}\frac{(1-e^{-\gamma (t^*+\Delta_0p)})}{(1-e^{-\gamma \Delta_0 p})}\frac{2}{\sqrt{\pi}} \ge 1
\end{equation}
which is possible after a time 
\begin{equation}
\label{1}
t^* = -\Delta_0 p\left[1  + \frac{1}{\gamma \Delta_0 p }\log \left(1 - \frac{\sqrt{(1+\sigma^2)}}{J_0(1-\e^{-\gamma\Delta_0(\theta+\theta')})}\frac{\sqrt{\pi}}{2}(1-e^{-\gamma \Delta_0 p})   \right)\right]\ .
\end{equation}
In order for this time to be finite $\theta$ should be larger than a certain threshold that is calculated in the main text. For time larger than $t^*$ we thus have that the opinions remain aligned with the signal even when this is removed, so the couplings $J$ can be calculated considering $g_i = \xi_i^\mu$ for the whole time interval $\Delta_0$. This means that to calculate $J$ at a time $t=N_p\Delta_0 p > t^*$ we need to add terms to the sum of integrals in Eq. \eqref{JA_nodrop}, which are integrals of the kind of Eq. \eqref{integral} albeit with the upper limit replaced by $a + \Delta_0$. This results in the following couplings:
\begin{eqnarray}
     J_{ij} &=&\nonumber \frac{J_0}{N} e^{-\gamma t}\left[ (e^{\gamma \Delta_0(\theta+\theta')}-1) \sum_{\mu=1}^{p}\xi_i^\mu\xi_j^\mu e^{(\mu-1)\Delta_0\gamma} \sum_{k=0}^{N_p^*-1}e^{\gamma \Delta_0 p k} \right. \\ 
     &\ & + \left. (e^{\gamma \Delta_0}-1) \sum_{\mu=1}^{p}\xi_i^\mu\xi_j^\mu e^{(\mu-1)\Delta_0\gamma} \sum_{k=N_p^*}^{N_p-1} e^{\gamma \Delta_0 p k}  \right] \\
     &=& \nonumber \frac{J_0}{N} e^{-\gamma t}\sum_{\mu=1}^{p}\xi_i^\mu\xi_j^\mu e^{(\mu-1)\Delta_0\gamma}\left[ (e^{\gamma \Delta_0(\theta+\theta')}-1)  \frac{(1-e^{\gamma\Delta_0p N_p^*})}{(1-e^{\gamma \Delta_0 p})} \right. \\ 
     &\ & + \left. (e^{\gamma \Delta_0}-1) \frac{(e^{\gamma\Delta_0p N_p^*}-e^{\gamma\Delta_0p N_p})}{(1-e^{\gamma \Delta_0 p})}  \right]
\end{eqnarray}
For simplicity in this appendix we calculated $J_{ij}$ only for discrete times multiple of $p\Delta_0$. However we can remark that it is possible to calculate them for any time, following the same reasoning used here.
\section{The role of signal strength for intermittent signals}
\label{appendix_I0}
	\begin{figure}
	\centering
	\includegraphics[scale=0.6]{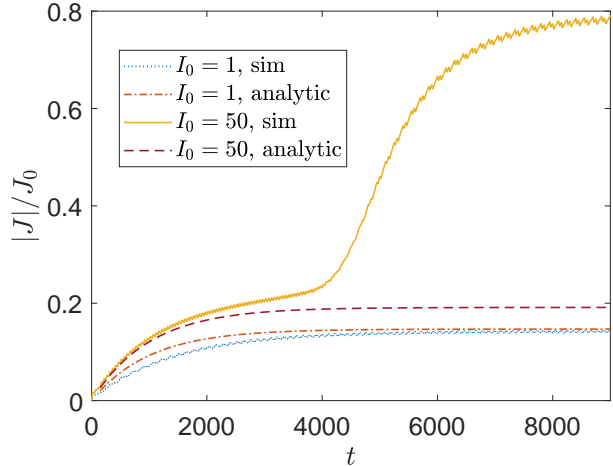}
	\caption{Intermittent signal dynamics. The figure shows an example of the evolution of  $|J|/J_0$ in time for $J$ analytical and from simulations and different signal strengths given three fixed patterns. The control parameters are $\gamma=10^{-3}, J_0=6, \Delta_0=50$ and $\Delta_1=37.5$.}
	\label{fig-appendix}
	\end{figure}
	The analytically predicted couplings $J$ do not always give the true norm $|J|$ of the couplings in our model, given that the approximations made in the calculations are of limited validity. In Fig. \ref{fig-appendix} we exhibit the dynamical evolution of $|J|/J_0$ for intermittent signals for a variety of signal strengths and observe that while the analytic theory does predict the true evolution reasonably well for small values $I_0$ of the signal strength, the analytical prediction fails qualitatively for very large values of $I_0$. The different curves in the figure correspond to dynamics with parameters other than the signal strength $I_0$ identical for all curves. The analytical prediction appears to work better for simulations with smaller $I_0$. This is due to approximations used in the estimation of the time $\theta'\Delta_0$ needed by the preference fields to decay to 0 when the signal is removed after a pattern presentation. Our estimation does in fact neglect the effect of the couplings during the decay.
 For $I_0=1$ the time $\theta'\Delta_0$ is not large enough to allow the couplings to grow much and neglecting their contribution in our calculation does not effect significantly the prediction of $|J|$. For $I_0=50$, the time $\Delta_0\theta'$ is large enough to allow the coupling to grow to sufficiently large values to permit eventual spontaneous recall, so neglecting them results in a substantial error in the prediction of $|J|$. In fact, in the case of $I_0=50$, after a time $t^*$ (approximately equal to the time of the steep increase of $|J|/J_0$ in the figure) the couplings have become sufficiently strong to sustain the opinion patterns when the signal is removed, kickstarting the positive feedback-loop which eventually results in the society being capable of nearly perfect spontaneous pattern retrieval --- a behaviour very different from the one predicted analytically for the given parameters (for which a finite $t^*$ does not exist). 

\bibliography{mybib}

\end{document}